\documentclass[aps,pre,twocolumn,groupedaddress,amsmath,showpacs]{revtex4-1}

\usepackage{graphicx}
\usepackage{bm}

\begin{document}



\title{Cracking Condition of Cohesionless Porous Materials in Drying Processes}



\author{So Kitsunezaki}
\email[]{kitsune@ki-rin.phys.nara-wu.ac.jp}
\homepage[]{http://www.complex.phys.nara-wu.ac.jp/~kitsune/main\_e.html}
\affiliation{Research Group of Physics, Division of Natural Sciences,
Faculty of Nara Women's University, Nara 630-8506, Japan}

\date{\today}

\begin{abstract}
The invasion of air into porous systems in drying processes is often
 localized in soft materials, such as colloidal
 suspensions and granular pastes, and it typically develops in the form
 of cracks before ordinary drying begins. 
To investigate such processes, we construct an invasion percolation
 model on a deformable lattice for cohesionless elastic systems, and with this model we derive the
 condition under which cracking occurs. 
A Griffith-like condition characterized by a dimensionless parameter is
 proposed, and its validity is checked numerically.
This condition indicates that the ease with which cracking occurs
 increases as the particles composing the material become smaller, as the
 rigidity of the system increases, and as the degree of heterogeneity
 characterizing the drying processes decreases.
\end{abstract}


\maketitle

\section{Introduction}

Contraction that results from drying often causes the formation of
cracks in pastelike materials such as wet granular materials and colloid suspensions. 
Compared with ordinary solids, such materials are soft in wet states due to weak
cohesion among constituent particles, and they possess unique properties characterizing the crack
formation that they exhibit, including memory effects, slow crack growth, and a diversity of crack patterns \cite{Pauchard03,Bohn05,Bohn05b,Nakahara06,Mal08,Pauchard09,Kitsunezaki09,Kitsunezaki10,Goehring10a,Goehring10b,Goehring13,Nakahara11}.
Considerable effort has been devoted to investigating the formation of crack
patterns in the contexts of physics, soil mechanics, and geology, while 
cracking conditions have been investigated
mainly in engineering applications.

During a drying process, the first crack in a paste generally appears in a capillary state,
 in which all pores (i.e., the spaces between solid particles) are filled with liquid,  
and the cracking process consists essentially of the invasion of air
 into these pores.
We believe that cracking conditions can be deduced directly from material
 properties related to drying, because on the microscopic level, cracking is indistinguishable
 from ordinary drying in the elementary processes of air invasion. 
However, drying and cracking have been treated separately in previous 
theories, although some studies have investigated the effect of large deformation
 and the order of ordinary drying and cracking in drying processes \cite{Routh99,Tirumkudulu05,Singh07,Russel08,Man08}. 

The main goal of this paper is to determine the Griffith criterion
associated with the drying process
of cohesionless porous systems.  
For this purpose, we focus on the limit of slow drying in a uniform elastic system.
The slow drying limit can be realized by decreasing the relative humidity $h$ 
quasistatically during the drying processes with fixed temperature $T$
and atmospheric pressure $P$.  
Under such conditions, both the water distribution and elastic
deformation of the material can be assumed to be in thermal
equilibrium or quasiequilibrium.
We also assume that the elastic relaxation is much faster than the 
redistribution of liquid, and we ignore plasticity, although it is
believed to be important for many type of pastes \cite{Wood90,Otsuki05,Nakahara06,Ooshida09,Kitsunezaki09,Kitsunezaki10,Nakahara11,Goehring13}.

To investigate the systems of interest, 
we extend an invasion percolation (IP) model to include elastic interactions.
It has been established that IP models are faithful models of drying
processes in porous materials, despite the great simplification they
employ of treating liquid distributions as binary distributions on a lattice
\cite{Wilkinson83,Wilkinson84,Wilkinson86,Du95,Yortsos97,Meakin00,Lovoll05,Yamazaki06,Nakanishi07}. 
Our extended model describes the cracklike invasion of air in soft
systems with large rigidity.
With it, we determine the condition for the formation of the first crack in terms of the free energy of the system. 
This condition is determined from the elastic properties of the material, the heterogeneity
characterizing the drying processes, and the size of particles forming
the paste. 

In this paper, 
we investigated a two-dimensional system corresponding to a
crosssection of a uniform layer of paste. 
We regard the bottom surface to be a fixed boundary and the top surface
to be a free boundary and assume that the evaporation of liquid occurs 
only from the top surface.
This paper is constructed as follows.
We propose a theoretical model based on thermodynamic considerations in
Sec.~\ref{Section:Slow Desiccation in Cohesionless Porous Material}. 
In Sec.~\ref{Section:Numerical
Simulations}, we report the results of numerical simulations using
this model.
We study the conditions governing the invasion of air and the formation
of the first crack in Sec.~\ref{Section:Invasion of air} and \ref{Section:Cracking},
respectively.
Conclusions are given in Sec.~\ref{Section:Conclusions}.

\section{Slow Desiccation in Cohesionless Porous Material}\label{Section:Slow Desiccation in Cohesionless Porous Material}

We regard the system as a mixture of solid particles and liquid forming
a paste layer, and the environment as air containing vapor, which exists
both inside and outside the paste layer. 
We treat this system and environment thermodynamically. 
In order to find the thermal equilibrium state for a
given $h$ with fixed $(T, P)$,    
we introduce a free energy 
$J\equiv F+P(V_s+V_l)-\mu_v(T,P,h)N_l$,
where $F$, $V_s$, $\mu_v$, $N_l$, and $V_l$ are the Helmholtz free
energy, the total volume of the constituent
particles, the chemical potential of the vapor, the number of liquid
molecules, and the volume of the liquid, respectively.  
We ignore the effect of gravity and assume that $V_s$ and $V_l/N_l$ are constant.

The Helmholtz free energy, $F$, is given by the sum $F\equiv
F_e+F_i+F_l$. 
Here, $F_e$ represents the deformation energy of the system, which
results from interactions among particles. 
The surface energy, $F_i$, is determined
by the liquid distribution and increases as the invasion of air
progresses,  
as shown in Appendix \ref{Equilibrium conditions}.
The free energy of the liquid, $F_l$, is given by $F_l=-P_lV_l+\mu_l
N_l$, as obtained from the Gibbs-D\"uhem relation, where $P_l$ is the
hydrostatic pressure of the liquid, and the chemical potential of the
liquid, $\mu_l$, is identical to
$\mu_v$. 
Substituting these forms into the above expression for $J$, and defining $p\equiv P-P_l$, we obtain 
\begin{equation}
J\equiv F_e+F_i+p(T,P,h)V_l+\mbox{const.} \label{J}
\end{equation}
The equilibrium state of the paste for given $h$ is determined by
minimizing $J$ with respect to the other state variables.  

The difference between the hydrostatic pressure of the fluid in the
pores and the atmospheric
pressure, $p$, is called the {\it negative pore pressure} in soil
mechanics. 
The pressure $p$ is uniform throughout the system in equilibrium states and 
determined uniquely by $h$ for given $T$ and $P$, in accordance with the
Kelvin equation, as indicated in Appendix \ref{Equilibrium conditions}.  
We adopt $p$ as the control parameter in place of $h$, because $p$ 
corresponds to the driving force of air invasion.
In the processes we consider, $p$ increases in time, while $h$ decreases
as a function of $p$.  

Here we give a brief remark concerning the minimization of $J$ for processes occurring at fixed $p$.
First, we note that even when $p$ increases quasistatically, 
a region into which air has invaded often expands abruptly, exhibiting
behavior similar to that in an avalanche process.  
However, if the redistribution and desiccation of liquid are sufficiently
slow, such a process can still be regarded as proceeding slowly, and,
therefore, we can assume that the system
is approximately in mechanical equilibrium throughout. 
While $V_l$ decreases gradually,   
instantaneous equilibrium states (characterized by instantaneous values
of $V_l$) can be obtained by minimizing $F$. 
Technically, this implies that we can always use the minimum principle of $J$ by     
including $V_l$ into the set of control parameters during such a process, because the
difference between $J$ and $F$, $P(V_s+V_l)-\mu_v(T,P,h(T,P,p)) N_l$, is constant 
if $V_l$ is fixed, in addition to $(T,P,p)$.  

\subsection{Rigid materials}
We use a regular lattice composed of $M$ cells each of volume
$\mathit{\Delta}V$ to represent the system.
The liquid distribution on the lattice is described by the set of
variables $\{\phi_1,\cdots,\phi_M\}$,  
where $\phi_m$ represents the ``dryness'' of the $m$th cell.
Each cell is assumed to be either wet ($\phi_m=0$) or dry
($\phi_m=1$), as in percolation models.
In order for such a treatment to be valid, we must assume that the cells 
possess microscopic volumes. 
We adopt the length of a cell as the unit of length in our model, which is
proportional to the linear extent of a particle, $r$, fixing all other
microscopic properties. 
In addition, we assume that all wet cells have the same liquid volume
fraction, $v_w$, and that no liquid exists in dry cells, for simplicity. 
The volume occupied by solid particles in each
cell is $(1-v_w)\mathit{\Delta}V$.

The total volume of liquid and surface energy are 
\begin{equation}
V_l\equiv \sum_{m=1}^M{\mathit\Delta
 V}v_w(1-\phi_m), \label{Vl}
\end{equation}
and
\begin{equation}
F_i\equiv \gamma_{la} \sum_{m=1}^M\mathit{\Delta}A_m\phi_{m}+\mbox{const.} \label{Fi},
\end{equation}
respectively, where $\gamma_{la}$ is the surface tension of the liquid-air interface. 
The quantity $\gamma_{la}\mathit{\Delta}A_m$ is the increase in the
surface energy needed to change the
state of the $m$th cell from wet to dry. 
This represents the resistance to drying.
In order to investigate heterogeneous material, we assume that
$\mathit{\Delta}A_m$ varies among the cells, with an average value $\overline{\mathit{\Delta}A}$. 
The characteristic pressure in a drying process,  
\begin{equation}
p_{\gamma}\equiv
 \gamma_{la}\frac{\overline{\mathit{\Delta}A}}{v_w\mathit{\Delta}V}\sim
 \frac{\gamma_{la}}{r}, 
\label{p_gamma}
\end{equation}
is typically on the order of the surface tension divided by $r$. 

Because $F_e$ is constant for rigid systems, we have 
\begin{equation}
J=\sum_{m=1}^M[\gamma_{la}\mathit{\Delta}A_m\phi_m+pv_w(1-\phi_m)\mathit{\Delta}V]+\mbox{const.} 
\label{J:rigid}
\end{equation}
Introducing the quantities $\tilde{J}\equiv J/v_wp_{\gamma}$,\ 
$\gamma_m \equiv \mathit{\Delta}A_m/\overline{\mathit{\Delta}A}$, and  
$\tilde{p}\equiv p/p_{\gamma}$, 
this expression can be written in the simplified form 
\begin{equation}
\tilde{J}=\sum_{m=1}^M\mathit{\Delta}V\{\gamma_m\phi_m+\tilde{p}(1-\phi_m)\},
\label{J:rigid:scaled}
\end{equation}
after removing the constant from $J$.

This free energy is minimized when the liquid distribution satisfies the
conditions 
\begin{equation}
\left\{\begin{array}{ll}
\phi_m=0 & \mbox{for}\ \gamma_m>\tilde{p}\\
\phi_m=1 & \mbox{for}\ \gamma_m\leq \tilde{p}\\
\end{array}
\right.. \label{wet/dry rule:rigid}
\end{equation}
If these conditions are applied only to cells adjacent to the dry-wet interface, this process corresponds to that described by the
conventional IP model without
a trapping rule~\cite{Wilkinson83,Wilkinson84,Wilkinson86}. 
In the present work, we assume that $\gamma_m$ is distributed uniformly
over the interval
	    $[1-\mathit{\Delta}\gamma,1+\mathit{\Delta}\gamma]$ and that
	    $\mathit{\Delta}\gamma \ll 1$. 
For large systems, air invades when $\tilde{p}\simeq 1-\mathit{\Delta}\gamma$ 
from the drying surface, and it percolates above a certain threshold
$\tilde{p}=p_c<1+\mathit{\Delta}\gamma$, as is well known.

Here, we note that the change of a state from wet to dry is reversible 
and may be caused simply 
by the redistribution of liquid, rather than desiccation, as observed by
L.~Xu {\it et al}. \cite{Xu08}. 
Also, desiccation in the vicinity of the free surface generally induces
liquid flow which causes the invasion of air far from this surface.
Such flows are often maintained in wet regions with high resistance to
drying and facilitate the drying processes \cite{Shokri13}.

\subsection{Extension to elastic materials}
We regard each cell as an elastic tile subject to uniform strains, in
order to describe the deformation of the system.
The elastic energy $F_e$ is determined by the strains as $F_e\{\bm{U}_m\}\equiv \sum_{m=1}^M\mathit{\Delta}V f_e(\bm{U}_m)$,
 where $\bm{U}_m\equiv (u^{(m)}_{\alpha\beta})$ is the strain tensor of the $m$th
	   cell and $f_e$ is the free energy density.
In order to simplify the situation, we make the following assumptions: 
(a1) all cells have the same elastic properties, 
(a2) $f_e$ does not depend on $\phi_m$, and 
(a3) $\mathit{\Delta}A_m$ does not depend on $\bm{U}_m$. 
Heterogeneity is introduced into the system only through the drying process. 

Coupling of $\phi_m$ and $\bm{U}_m$ is introduced through $V_l$. 
The dilation of a cell results from the influx of liquid for a wet cell
and an influx of air for a dry cell. 
To account for such phenomena, we add the volumetric strain 
$u_{\alpha\alpha}^{(m)}$ to $v_w$ in Eq.~(\ref{Vl}).
This yields the free energy 
\begin{eqnarray}
J&=&F_e\{\bm{U}_m\}+
\sum_{m=1}^M[\gamma_{la}\mathit{\Delta}A_m\phi_m
\nonumber \\
& & +p(v_w+u_{\alpha\alpha}^{(m)})(1-\phi_m)\mathit{\Delta}V]
+\mbox{const}, \label{J:elastic}
\end{eqnarray}
where Einstein's summation rule is applied to repeated Greek indices. 
Then, introducing $\tilde{\bm{U}}_m\equiv \bm{U}_m/v_w$ and
$\tilde{f_e}(\tilde{\bm{U}}_m)\equiv f_e(\bm{U}_m)/v_wp_{\gamma}$, we
obtain 
\begin{equation}
\tilde{J}=\sum_{m=1}^M\mathit{\Delta}V\{\tilde{f}_e(\tilde{\bm{U}}_m)+\gamma_{m}\phi_m+\tilde{p}(1+\tilde{u}^{(m)}_{\alpha\alpha})(1-\phi_m)\}.
\label{J:elastic:scaled}
\end{equation}
We note that the assumption (a3) can be weakened slightly in the case that
$\mathit{\Delta}A_m$ depends on $\bm{U}_m$ linearly as
$\mathit{\Delta}A_m=\mathit{\Delta}\overline{A}(\gamma_m+\gamma'\tilde{u}_{\alpha\alpha}^{(m)})$, 
where $\gamma'$ is a constant, because 
an expression identical to Eq.~(\ref{J:elastic:scaled}) can be  
 obtained through an appropriate transformation of the variables.

With the elastic energy included, the conditions to determine the liquid distribution
	   are revised from those appearing in Eq.~(\ref{wet/dry rule:rigid}) to 
\begin{equation}
\left\{\begin{array}{ll}
\phi_m=0 & \mbox{for}\ \gamma_m>\tilde{p}(1+\tilde{u}^{(m)}_{\alpha\alpha})\\
\phi_m=1 & \mbox{for}\ \gamma_m\leq \tilde{p}(1+\tilde{u}^{(m)}_{\alpha\alpha})\\
\end{array}
\right.. \label{wet/dry rule:elastic}
\end{equation}
These conditions imply that the resistance to drying decreases as a cell
expands, because the expansion of a cell decreases the cost in surface energy 
required to remove a unit volume of liquid. 

The free energy has a minimum with respect to $\{\tilde{\bm{U}}_m\}$ in the equilibrium state. 
Minimizing Eq.~(\ref{J:elastic:scaled}) in the continuum limit  
gives the stress balance equation 
\begin{equation}
\frac{\partial \tilde{\sigma}_{\alpha\beta}}{\partial x_{\beta}}=0,
\label{stress balance equation}
\end{equation}
where $x_{\beta}$ represents the space coordinates and  
the stresses $\tilde{\sigma}_{\alpha\beta}$ are defined by 
\begin{equation}
\tilde{\sigma}_{\alpha\beta}\equiv \frac{\partial \tilde{f}_e}{\partial
 \tilde{u}_{\alpha\beta}}+\tilde{p}(1-\phi)\delta_{\alpha\beta}.
\end{equation}
Solid particles in wet regions are subject to compressive pressure $\tilde{p}$
	   from both the free surface and the interface with dry regions.

\subsection{Elastic energy of cohesionless materials}
Soft materials that exhibit drying cracks, such as colloid suspensions and
wet granular materials,  generally have nonlinear elastic
properties. 
In many cases, the cohesive interactions of the constituent 
particles are very weak in comparison with excluded volume interactions, 
and materials in capillary states hold their shape under compressive
stresses caused by negative pore pressures.
\footnote{
Strong cohesion is often formed among
constituent particles, typically after drying~\cite{Kendall87}. 
In such cases, cracking has common properties with typical brittle 
fracture in contrast to that in capillary states~\cite{Kitsunezaki09}.}
The elastic energy and moduli practically vanish unless the system is
subject to compressive stresses.  
Therefore, we assume that $\tilde{f}_e(\tilde{\bm{U}})=0$ for
$\tilde{u}_{\alpha\alpha}\geq 0$  
and that the elastic moduli vanish for $\tilde{u}_{\alpha\alpha}=0$.  

We need to choose an appropriate function of $\tilde{f}_e(\tilde{\bm{U}})$
for $\tilde{u}_{\alpha\alpha}<0$, because there is no established
general constituent relation. 
If we assume an isotropic analytic function for the elastic moduli,
$\tilde{f}_e$ can be approximated in the form of a third-order elasticity as  
$\tilde{f}_e=-\lambda \tilde{u}_{\alpha\alpha}^3-\mu \tilde{u}_{\xi\xi}\tilde{u}_{\alpha\beta}^2$ 
for small $\tilde{u}_{\alpha\alpha}$, 
where $\lambda$ and $\mu$ are positive constants. 
L. Goehring reported that this third-order elasticity accurately describes 
the results of compression tests with cornstarch paste \cite{Goehring09}. 
Another choice is 
$\tilde{f}_e=\sqrt{-\tilde{u}_{\xi\xi}}(\lambda
\tilde{u}_{\alpha\alpha}^2+\mu \tilde{u}_{\alpha\beta}^2)$, 
which is obtained theoretically by assuming Hertzian contacts among particles and affine
deformation \cite{Jiang07,Russel08}. 

We assume the following general homogeneous form including these
choices:   
\begin{equation}
\tilde{f}_e(\tilde{\bm{U}})=g(\tilde{u}_{\xi\xi})\left[\frac{1}{2}\tilde{K}
 \tilde{u}_{\alpha\alpha}^2+\tilde{G}\left(\tilde{u}_{\alpha\beta}-\frac{1}{2}\tilde{u}_{\eta\eta}\delta_{\alpha\beta}\right)^2\right], 
\label{fe}
\end{equation}
where $\tilde{K}$ and $\tilde{G}$ are positive constants. 
The function $g(x)$ takes the power-law form 
\begin{equation}
g(x)=\left\{\begin{array}{ll}
(-x)^{\nu-1} & x<0 \\
0            & x\geq 0 
\end{array}
\right.
\label{g:homogenerous},
\end{equation}
where $\nu=2$ for third-order elasticity and $\nu=3/2$ in the case of 
Hertzian contacts.
The bulk modulus and rigidity depend on the state of the material, due
to the nonlinearity.
As shown in Appendix \ref{Isotropic compressive states}, they are
proportional to $g(\tilde{u}_{\alpha\alpha})$ for isotropic 
compressive states,
 and $2\tilde{G}<\nu(\nu+1)\tilde{K}$ is required for most materials with
 positive Poisson's ratio. 
We investigated our model mainly for large $\tilde{K}$, as 
Eq.~(\ref{fe}) generally holds for small deformations.

\section{Numerical Simulations}\label{Section:Numerical Simulations}

\subsection{Methods}
\begin{figure}
\includegraphics[width=6cm]{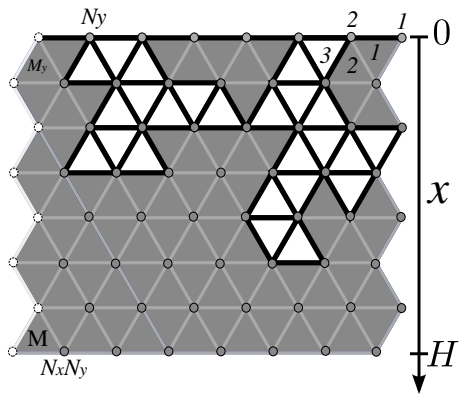}
\caption{Triangular lattice with $N_x\times N_y$ vertices and $M=2N_y(N_x-1)$ cells. 
The $x$-axis is normal to the surfaces of the system, with the value of
 $x$ representing the distance from the top surface. 
Dry cells are enclosed by bold lines.
}
\label{lattice}
\end{figure}

\begin{figure}
\includegraphics[width=8.5cm]{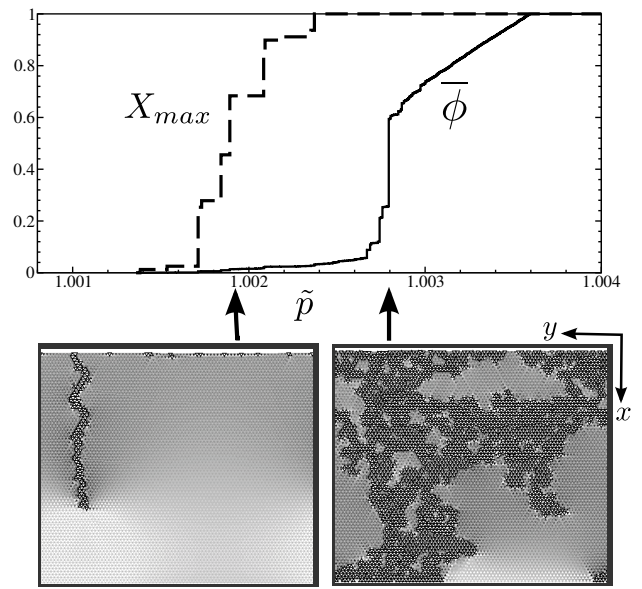}
\caption{The drying process following the cracklike invasion of air.  
The graph displays the results for the dry fraction $\overline{\phi}$ and the
 maximum depth of a dry cell, $X_{max}$, as functions of $\tilde{p}$ 
obtained from a numerical simulation with
 $(\tilde{K},\tilde{G})=(10^5,2\times 10^4)$,
 $\mathit{\Delta}\gamma=0.001$, and $N_x=N_y=80$.  
The bottom snapshots correspond to
 $(\tilde{p}-1)/\mathit{\Delta}\gamma=1.19$ and $2.79$. 
}
\label{Drying process}
\end{figure}

We consider a uniform layer from whose top surface (which is a
one-dimensional interface) liquid is desiccated. 
We carried out numerical simulations using third-order elasticity
($\nu=2$) on a two-dimensional triangular lattice, as shown in
Fig.~\ref{lattice}.
The lattice has $N_x\times N_y$ vertices, and the layer
thickness is $H\equiv \sqrt{3}(N_x-1)/2$.

In order to avoid erroneous numerical convergence due to the singularity at
$\tilde{u}_{\alpha\alpha}=0$ for $g'(x)$, 
we used the smooth function
\begin{equation}
g(x)=\left\{\begin{array}{ll}
-x & \mbox{for}\ b x< -1 \\
\frac{1}{4b}(b x-1)^2 & \mbox{for}\ |bx|\leq 1 \\
0 &  \mbox{for}\ bx>1 \\
\end{array}
\right.,\label{g:numerical}
\end{equation} 
with a large positive constant $b=10^4$, instead of
the form given in Eq.~(\ref{g:homogenerous}). 

The numerical method we used is essentially the same as that used in
Ref.~\cite{Kitsunezaki99}.
The deformation of the lattice is described by the displacements of the
vertices, $\{\bm{u}_1,\cdots,\bm{u}_{N}\}$ ($N\equiv N_xN_y$), 
which determine $\{\tilde{\bm{U}}_m\}$. 
The top surface is a free boundary in contact with air. 
The interface between air and wet cells is treated as the dry-wet interface.
The bottom surface is a fixed boundary with respect to $\{\bm{u}_n\}$ and a reflecting boundary
 with respect to $\{\phi_m\}$. 
Periodic boundary conditions are used along the $y$ direction. 
The following procedures were repeated in the numerical simulations from  
the initial conditions in which all cells were wet and undeformed.  
\begin{description}
\item[(P1)] The displacements $\{\bm{u}_n\}$ were calculated by minimizing
	    $\tilde{J}$ for fixed $\{\phi_m\}$ and $\tilde{p}$ in Eqs.~ (\ref{J:elastic:scaled}) and (\ref{fe})
	   using the conjugate gradient method
\cite{Press02}.
\item[(P2)] The conditions (\ref{wet/dry rule:elastic}) were checked for all cells
contacting the dry-wet interface. 
If these conditions were satisfied, we increased $\tilde{p}$ by $\mathit{\Delta}\tilde{p}$ and then returned to (P1). 
\item[(P3)] If the $m$th cell did not satisfy these conditions, we changed $\phi_m$ 
	   from wet (dry) to dry (wet) and returned to (P1). 
\end{description}
In (P3), if we found more than one cell that did not satisfy the
conditions (\ref{wet/dry rule:elastic}), 
we changed only the state of the most unstable cell, i.e., that with the
largest deviation from the condition $\gamma_m=\tilde{p}(1+\tilde{u}_{\alpha\alpha}^{(m)})$.  
Under this procedure, $V_l$ changes gradually for fixed $\tilde{p}$. 
In the simulations whose results are presented here, we used  
$\mathit{\Delta}\tilde{p}=\mathit{\Delta}\gamma/500$, and the tolerance
of the conjugate gradient method was $3\times 10^{-11}$.
The numerical results were confirmed to be essentially the same with
those obtained for smaller $\mathit{\Delta}\tilde{p}$ and tolerance.

\subsection{Results}
\begin{figure*}
\includegraphics[width=16cm]{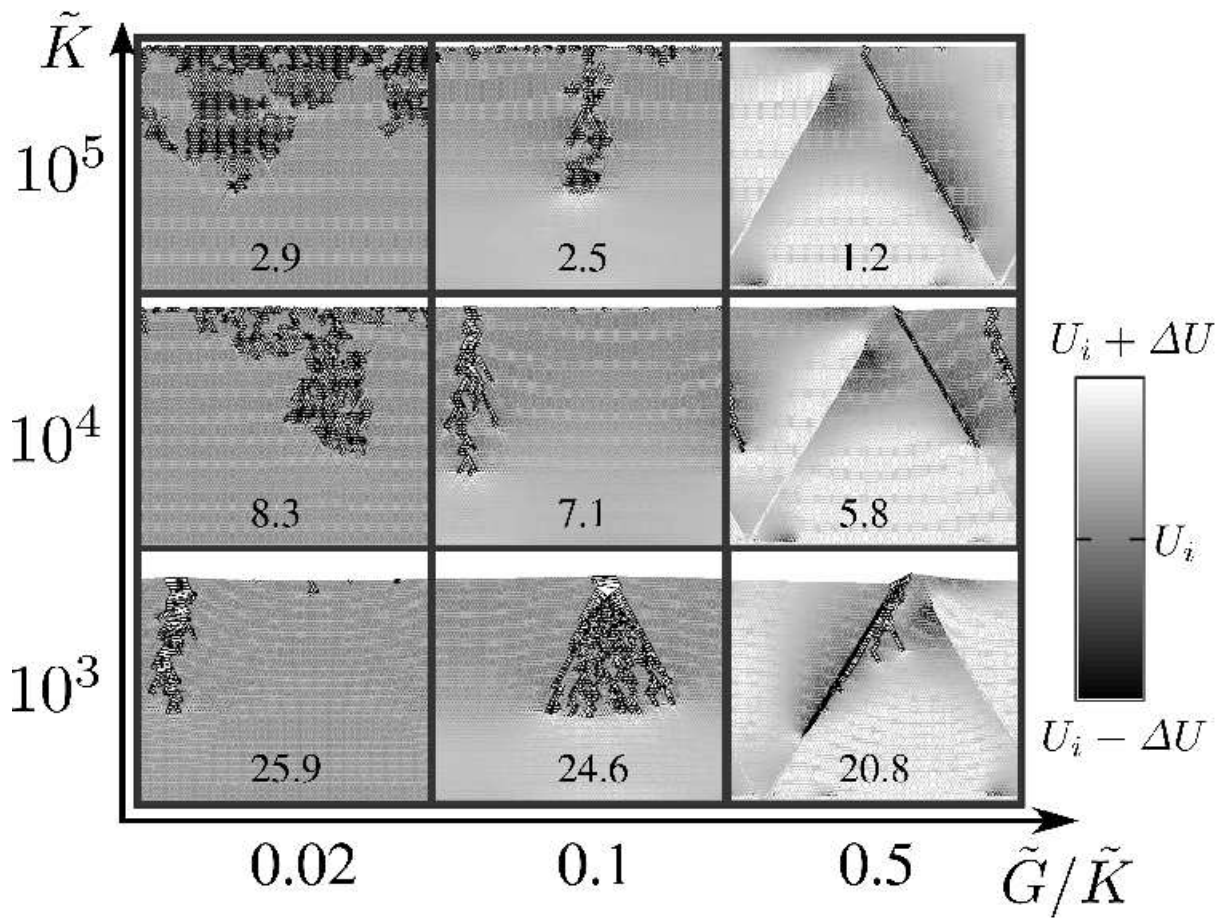}
\caption{Snapshots of air invasion for $\mathit{\Delta}\gamma=0.001$, 
$\tilde{K}=10^3, 10^4, 10^5$ and $\tilde{G}/\tilde{K}=0.02,0.1,0.5$. 
In all cases, $N_x=N_y=80$. 
The number appearing in each figure indicates the value of $(\tilde{p}-1)/\mathit{\Delta}\gamma$
 for that snapshot.  
The deformation of the lattice is scaled by a factor of $5$ for viewability, and 
the volumetric strain $\tilde{u}_{\alpha\alpha}$ is indicated by
 the gray scale, which corresponds to the 
 interval $[\tilde{U}_i-\mathit{\Delta}U,\tilde{U}_i+\mathit{\Delta}U]$, where $\tilde{U}_i$ is
 given by Eq.~(\ref{Ui}) and $\mathit{\Delta}U=1/(6\sqrt{10\tilde{K}})$. 
}
\label{FIG:KG}
\end{figure*}

Our numerical results indicate that the cracklike invasion of air
occurs readily in soft systems with large rigidity and small heterogeneity.

Figure \ref{Drying process} depicts a typical process of cracklike invasion and
subsequent drying. 
As soon as a cell on the top surface dries,  
air penetrates rapidly into the bulk and this results in the formation
of a one-dimensional dry region.
This dry region expands gradually from both the top surface and the crack
 line as $\tilde{p}$ increases. 
The graph in this figure plots the fraction of the entire system in the
dry region, 
$\overline{\phi}\equiv \sum_{m=1}^M \phi_m/M$,
 and the maximum depth of a dry cell divided by the thickness, 
$X_{max}\equiv \max_{m}{\{x_m| \phi_m=1\}}/H$, 
as functions of $\tilde{p}$. 
It is found that soft material resists the invasion of air through shrinkage. 
When shrinkage occurs, the pressure required for air invasion to begin 
is larger than that in the case of the conventional IP model, $\tilde{p}\simeq 1-\mathit{\Delta}\gamma$.
On the other hand, the percolation threshold 
decreases drastically if the cracklike invasion of air occurs.

Figure \ref{FIG:KG} displays typical snapshots for nine sets of $(\tilde{K},\tilde{G})$ and 
fixed $\mathit{\Delta}\gamma$. 
These were taken after a dry region had developed beyond a depth of
approximately half of the total depth. 
For sufficiently large $\tilde{K}$ or small $\tilde{G}$, the air invasion process occurs in the same
manner as for the conventional IP model. 
It always occurs at $\tilde{G}=0$ for any value of $\tilde{K}$, as described below.  
The dry region expands intermittently and gradually as the pressure $\tilde{p}$ increases. 
Contrastingly, for large $\tilde{G}$ and small $\tilde{K}$, cracklike
air invasion occurs first.
This dry region essentially corresponds to a mode I crack, because the
cells in this region are expanded (i.e., $\tilde{u}_{\alpha\alpha}>0$), while the surrounding wet region shrinks. 
For sufficiently large $\tilde{G}$ and small $\tilde{K}$, however, shear
bands (mode II cracks) often form ahead of cracks in the wet region.  
The directions of the shear bands reflect the anisotropy of the
triangular lattice.
The wet cells in such shear bands expand until $\tilde{u}_{\alpha\alpha}\simeq
0$, and cracklike air invasion develops along shear bands.
Shear bands sometimes appear and disappear during the invasion of air.

\begin{figure*}
\begin{center}
\includegraphics[width=14cm]{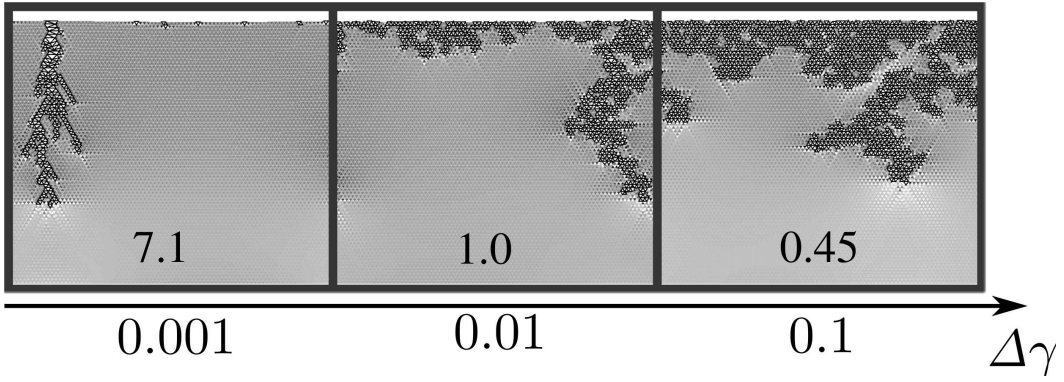}
\caption{Snapshots of air invasions for $\mathit{\Delta}\gamma=0.001, 0.01,
 0.1$ and fixed $(\tilde{K},\tilde{G})=(10^4,10^3)$. 
These snapshots are displayed in the same manner as those in Fig.~\ref{FIG:KG}. 
In all cases, $N_x=N_y=80$. 
The number in each figure indicates
the value of $(\tilde{p}-1)/\mathit{\Delta}\gamma$ for the respective
 values of $\mathit{\Delta}\gamma$.
}
\label{FIG:dgam}
\end{center}
\end{figure*}

Figure \ref{FIG:dgam} elucidates the dependence of the heterogeneity for fixed $(\tilde{K},\tilde{G})$. 
As $\mathit{\Delta}\gamma$ increases, the air invasion process changes from cracklike to conventional IP-like. 
The value of $\mathit{\Delta}\gamma$ at which this change occurs depends on $(\tilde{K},\tilde{G})$, as
discussed in Sec.~\ref{Section:Cracking}.

The transition between cracklike and conventional IP-like invasion
has been reported for some heterogeneous systems with long-range interactions~\cite{Chakrabarti97,Shekhawat12}. 
In particular, R.~Holtzman {\it et al}. investigated the displacement of
immiscible fluid in preloaded granular systems 
and reported three types of invasion: fracturing, capillary fingering (CF) 
, and viscous fingering(VF)~\cite{Holtzman10,Holtzman12}. 
Fracturing and CF in their systems correspond to cracklike and conventional IP-like 
invasions, respectively, although loading arises from increasing negative pore
pressures in our systems. 
VF is caused by the effect of pressure gradient. 
Although it does not occur in the slow drying limit,   
the gradient of negative pore pressures becomes important for fast drying and 
results in directional cracking generally~\cite{Kitsunezaki11, Goehring13}.

\section{Invasion of air}\label{Section:Invasion of air}

When the invasion of air does not occur yet, the system is uniform and contracts only along
the $x$ direction as $\tilde{p}$ increases with drying. 
Such uniaxial compressive states are described by $\phi=0$ and 
\begin{equation}
(\tilde{u}_{\alpha\beta})=\left(\begin{array}{cc}
 \tilde{U}_a & 0 \\
 0 & 0 
\end{array}\right)\equiv \tilde{\bm{U}}_a,
\label{uniaxial compressive states}
\end{equation}
where $\tilde{U}_a$ is determined from the free boundary condition $\tilde{\sigma}_{xx}=\partial \tilde{f}_e/\partial
 \tilde{U}_a+\tilde{p}=0$ on the top line. 
Substituting these values into Eqs.~(\ref{fe}) and
 (\ref{g:homogenerous}), we obtain 
$\tilde{f}_e=(\tilde{K}+\tilde{G})(-\tilde{U}_a)^{\nu+1}/2$ and then 
\begin{equation}
\tilde{U}_a=-\left\{\frac{2\tilde{p}}{(\nu+1)(\tilde{K}+\tilde{G})}\right\}^{\frac{1}{\nu}}
\ \mbox{and}\ 
\tilde{f}_e(\tilde{\bm{U}}_a)=-\frac{1}{\nu+1}\tilde{p}\tilde{U}_a.
\label{Ua}
\end{equation}

If the top surface is sufficiently long, the invasion of air begins when
the second condition in (\ref{wet/dry rule:elastic}) holds at a cell on
the surface for which we have 
\begin{equation}
1-\mathit{\Delta}\gamma=\tilde{p}(1+\tilde{U}_a), 
\label{air invasion condition}
\end{equation}
because the smallest $\gamma_m$ is $1-\mathit{\Delta}\gamma$.
The pressure at which the invasion of air begins is determined by this
 condition and increases as $\tilde{K}+\tilde{G}$ decreases. 
Because the right-hand side of Eq.~(\ref{air invasion condition}) has a maximum at
$\tilde{U}_a=-\nu/(\nu+1)$, the invasion of air never occurs for systems
so soft that the condition
\begin{equation}
\tilde{K}+\tilde{G}<
 2\left(1+\frac{1}{\nu}\right)^{\nu}(1-\mathit{\Delta}\gamma)\simeq
 2\left(1+\frac{1}{\nu}\right)^{\nu}
\label{hardness condition for air invasion}
\end{equation}
holds. This case corresponds to wet sintering in this system~\cite{Routh99,Pauchard03,Tirumkudulu05,Singh07,Russel08}.

\

Let us first consider a drying process for $\tilde{G}=0$, that is, the
case with no rigidity. 
In this case, the system does not exhibit
cracking and dries in the same manner as in the conventional IP model.  
We can solve the stress balance equation (\ref{stress balance equation})
easily in the case $\tilde{G}=0$, and we obtain isotropic stress states
with 
$\tilde{\sigma}_{\alpha\beta}=0$ 
and 
\begin{equation}
\tilde{u}_{\alpha\alpha}=-\left\{\frac{2\tilde{p}}{(\nu+1)\tilde{K}}\right\}^{\frac{1}{\nu}}\equiv
 \tilde{U}_i
\label{Ui}
\end{equation}
for all wet cells and $\tilde{u}_{\alpha\alpha}\geq 0$ for dry cells.
The elastic energy of a wet cell is the same as that of a cell in the
isotropic compressive state, 
\begin{equation}
 \tilde{f}_e(\tilde{\bm{U}}_i)=-\frac{1}{\nu+1}\tilde{p}\tilde{U}_i,
\ \mbox{where }\ 
\tilde{\bm{U}}_i\equiv \frac{\tilde{U}_i}{2}\left(\begin{array}{cc}
 1 & 0 \\
 0 & 1
\end{array}\right).
\label{Ui:fe}
\end{equation}
The pressure at which the $m$th cell is allowed to dry is determined 
 from $\gamma_m=\tilde{p}[1+\tilde{U}_i(\tilde{p})]$, in accordance with
 the conditions in (\ref{wet/dry rule:elastic}), along with the condition $\tilde{U}_i>-\nu/(\nu+1)$.
Because the pressure always increases monotonically as $\gamma_m$ increases,  
the order of drying does not depend on $\tilde{K}$.

The above conclusion concerning the nature of the drying can also be obtained
by considering the free energy, $J$.
To show this, we first note that additional energy is 
required to dry cells on an elastic lattice, because the elastic energy stored in a
 cell is dissipated quickly through dilation following drying. 
In order to develop a dry region $D$ for a fixed value of $\tilde{p}$, 
the amount by which the free energy $\tilde{J}$ decreases must be
larger than the dissipation,  
\begin{equation}
R\equiv \sum_{m\in D}\mathit{\Delta}V\tilde{f}_e(\tilde{\bm{U}}_m^{(\mathrm{dry})}),
\label{R}
\end{equation}
where $\tilde{\bm{U}}_m^{(\mathrm{dry})}$ is the strain on the $m$th
cell at the time that it changes from wet to dry.
This condition can be written  
\begin{equation}
\tilde{J}(\tilde{p};\emptyset)-\tilde{J}(\tilde{p};D)-R=\sum_{m\in
 D}\mathit{\Delta}V\left(\tilde{p}(1+\tilde{U}_i)-\gamma_{m}\right)>0,   
\end{equation}
because $f_e(\tilde{\bm{U}}_m^{(\mathrm{dry})})=f_e(\tilde{\bm{U}}_i)$,
and Eq.~(\ref{J:elastic:scaled}) can be rewritten as $\tilde{J}(\tilde{p};D)
=\sum_{m\notin
 D}\mathit{\Delta}V\left\{\tilde{f}_e(\tilde{\bm{U}}_i)+\tilde{p}(1+\tilde{U}_i)\right\}+\sum_{m\in
 D}\mathit{\Delta}V\gamma_{m}$, where $D=\emptyset$ (empty set) corresponds to the initial state, in which all cells are wet.
When the invasion of air begins with the condition (\ref{air invasion
condition}), the right-hand side is negative, because $\tilde{p}(1+\tilde{U}_i)=1-\mathit{\Delta}\gamma<\gamma_m$.
Therefore, the dry region can expand only in a step-by-step manner with
increasing $\tilde{p}$, and, hence, there is no cracking.

\section{Cracking}\label{Section:Cracking}

\subsection{Fracture criterion for the first crack}
Next, we consider the case $\tilde{G}\neq 0$ and investigate the
cracking condition.  
Cracking in our system consists of a quasi-one-dimensional invasion of
air. 
It is facilitated by the release of energy from a surrounding wet region.
Just as in the situation considered in the previous section, a crack
develops if the amount by which the free energy $\tilde{J}$ decreases is
larger than the dissipation $R$ when the invasion of air begins. 

If there appears a crack of length $L$ that is perpendicular to the free
surface in a sufficiently large system with fixed $\tilde{p}$, 
this crack forms an air-invaded region $D$ and 
it causes the elastic energy stored in $D$ to dissipate locally. 
The wet region $S$ surrounding the crack, from which the elastic energy is
released, has an area of approximately $L\times L$, because $L$ is the only characteristic length in this system. 
The states in this region change from the uniaxial compressive state,
characterized by $\tilde{\bm{U}}_a$, to an approximately isotropic
compressive state, characterized by $\tilde{\bm{U}}_i$. 
Measuring $L$ by the number of cells in $D$, the decrease of $\tilde{J}$ is estimated as 
\begin{eqnarray}
\tilde{J}(\tilde{p};\emptyset)&-&\tilde{J}(\tilde{p};D)
\nonumber \\
& \simeq &
L^2\mathit{\Delta}V\left\{\tilde{f}_e(\tilde{\bm{U}}_a)-\tilde{f}_e(\tilde{\bm{U}}_i)+\tilde{p}(\tilde{U}_a-\tilde{U}_i)\right\}
\nonumber \\
&+&L\mathit{\Delta}V\left\{\tilde{f}_e(\tilde{\bm{U}}_a)+\tilde{p}(1+\tilde{U}_a)-\gamma_D(L)\right\},
\label{Decrease of J for cracking}
\end{eqnarray}
where the average of $\gamma_m$ in $D$ is represented by 
\begin{equation}
\gamma_D(L)\equiv \frac{1}{L}\sum_{m\in D}\gamma_m.
\end{equation}

The elastic energy of the $m$th cell in $D$, $f_e(\tilde{\bm{U}}_m)$, dissipates with drying at $\gamma_m\simeq \tilde{p}(1+\tilde{u}_{\alpha\alpha}^{(m)})$.
A cell is deformed at a crack tip and expands by at least an amount 
 $\tilde{u}_{\alpha\alpha}^{(m)}-\tilde{U}_a\simeq
 (\gamma_m-1+\mathit{\Delta}\gamma)/\tilde{p}=O(\mathit{\Delta}\gamma)$
 in comparison with Eq.~(\ref{air invasion condition}).
However, as shown in Appendix \ref{Perturbation from a uniaxial compressive
 state}, 
the energy in this case is the same as $f_e(\tilde{\bm{U}}_a)$ to first order in $\mathit{\Delta}\gamma$, because
the cell contacts a dry-wet interface.
Thus, the minimum dissipation for air invasion is 
\begin{equation}
R=L\mathit{\Delta}V\tilde{f}_e(\tilde{\bm{U}}_a)+O((\mathit{\Delta}\gamma)^2).
\label{R:cracking}
\end{equation}
The cracking condition
$\tilde{J}(\tilde{p};\emptyset)-\tilde{J}(\tilde{p};D)-R>0$ for fixed
$\tilde{p}$ and $D$ is approximated as 
\begin{eqnarray}
L\left\{\tilde{f}_e(\tilde{\bm{U}}_a) - \tilde{f}_e(\tilde{\bm{U}}_i)+\tilde{p}(\tilde{U}_a-\tilde{U}_i)\right\}
\hspace{5em} \nonumber \\
\gtrsim
 \gamma_D(L)-\tilde{p}(1+\tilde{U}_a)=\gamma_D(L)-1+\mathit{\Delta}\gamma,
\label{cracking condition for all cases}
\end{eqnarray}
where we have used Eq.~(\ref{air invasion condition}). 
The right-hand side of this expression is positive and order $\mathit{\Delta}\gamma$. 
The cracking condition of the system is determined by finding the
minimum value of $\gamma_D(L)$ over all possible crack paths $D$.
This yields 
\begin{equation}
L\left\{\tilde{f}_e(\tilde{\bm{U}}_a)-\tilde{f}_e(\tilde{\bm{U}}_i)+\tilde{p}(\tilde{U}_a-\tilde{U}_i)\right\}
\gtrsim c(L)\mathit{\Delta}\gamma,
\label{cracking condition}
\end{equation}
where $c(L)\mathit{\Delta}\gamma\equiv
\min_{D}{(\gamma_D(L)-1+\mathit{\Delta}\gamma)}$.
The quantity $c(L)$ increases from $c(1)\simeq 0$ to a value less than $1$ as $L$ increases. 

The condition derived above corresponds to the Griffith criterion for the first crack
in a drying process. 
It indicates that   
there is a critical crack length $L=L_c$ beyond which unstable crack
growth occurs. 
The left-hand side of Eq.~(\ref{cracking condition}) is the energy
released when the crack advances by one cell.  
The right-hand side, $c(L)\mathit{\Delta}\gamma$, is the additional energy required to dry a
cell at a crack tip.  
The critical length, $L_c$, is determined by the nondimensional parameter 
\begin{equation}
\Gamma\equiv
 \mathit{\Delta}\gamma\left\{\tilde{f}_e({\bm
 U}_a)-\tilde{f}_e(\tilde{\bm{U}}_i)+\tilde{p}(\tilde{U}_a-\tilde{U}_i)\right\}^{-1}
\label{Definition of gamma}
\end{equation}
as $L_c=c(L_c)\Gamma$.
Substituting Eqs.~(\ref{Ua}), (\ref{Ui}) and (\ref{Ui:fe}) into
Eq.~(\ref{Definition of gamma}), we obtain 
\begin{equation}
\Gamma
=\left(1+\frac{1}{\nu}\right)\frac{\mathit{\Delta}\gamma}{\tilde{p}(\tilde{U}_a-\tilde{U}_i)}
=\frac{\tilde{K}^{\frac{1}{\nu}}\mathit{\Delta}\gamma}{g_{\nu}\left(\frac{\tilde{G}}{\tilde{K}}\right)\tilde{p}^{1+\frac{1}{\nu}}},
\label{Gamma}
\end{equation}
where 
\begin{equation}
g_{\nu}(x)\equiv
 \left(\frac{2}{\nu+1}\right)^{1+\frac{1}{\nu}}\frac{\nu}{2}\left\{1-(1+x)^{-\frac{1}{\nu}}\right\}. 
\end{equation}
The quantity $g_{\nu}(x)$ is an increasing function of $x$ and is
approximately proportional to $x/2$ for $x\ll 1$.
From Eq.~(\ref{air invasion condition}), we find $\tilde{p}\simeq 1$ for $\tilde{K}\gg 1$ and
$\mathit{\Delta}\gamma\ll 1$.

\subsection{Verification in numerical simulations}
\begin{figure}
\includegraphics[width=7.0cm]{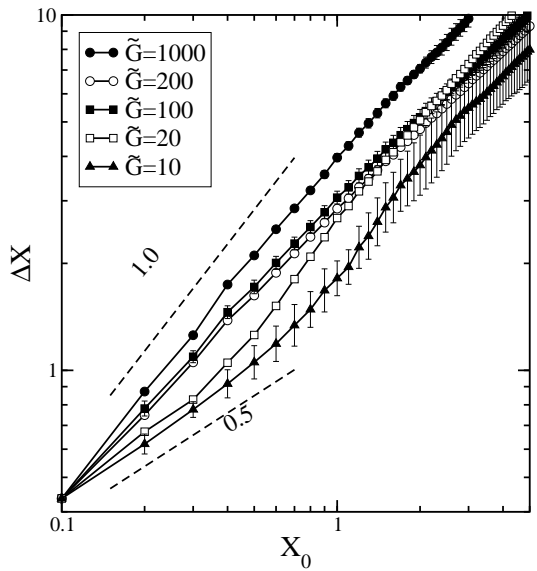} 
\caption{Dependence of $\mathit{\Delta}X$ on $X_0$ for $\tilde{K}=1000$ and
 $\mathit{\Delta}\gamma=0.001$. 
The results for each value of $\tilde{G}$ were calculated from 
 the average of $X_1$ over eight numerical simulations on a lattice 
 with $N_x=N_y=40$.  
The error bars for the plots with $\tilde{G}=20$ and $200$ are omitted for clarity.}
\label{DeltaX}
\end{figure}

\begin{figure}
\includegraphics[width=7.0cm]{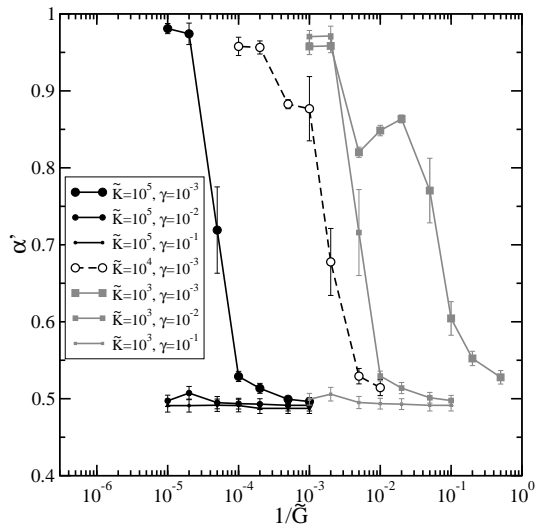} 
 \caption{Dependence of $\alpha'$ on $1/\tilde{G}$ for various $\tilde{K}$ and
 $\mathit{\Delta}\gamma$. 
These results were obtained by applying a least-square method to a log-log plot of $\mathit\Delta{X}$ and
 $X_0$ for $X_0<1$ to calculate $\alpha'$.}
\label{alpha-1/G} 
\end{figure}

In order to verify the criterion appearing in Eq.~(\ref{cracking
condition}) on the basis of our numerical simulations, it is necessary
to quantify the transition from the invasion of the conventional IP
model to cracklike invasion in the initial stage.  
We calculated the quantities 
\begin{equation}
X_k\equiv  \frac{1}{N_y}\sum_{m=1}^M x_m^k\phi_m 
\end{equation}
for $k=0$ and $1$ and investigated the dependence of $\mathit{\Delta}X\equiv \sqrt{2X_1-X_0^2}$ on $X_0$. 

The conventional IP model is characterized by the self-affine growth of a dry-wet interface. 
Assuming the interface to be a single-valued function $x=h(y)$ in the initial stages, the standard deviation $\sqrt{\langle(h-\langle h\rangle)^2\rangle}$ increases as a power function of the average height
 $\langle h\rangle$, 
where $\langle \cdot \rangle$ represents the average over $y$.
Because $\phi_m=1$ for $0\leq x_m<h(y)$, 
the standard deviation corresponds to $\mathit{\Delta}X$, as we have 
$X_1\simeq \langle \int_0^h dx x\rangle=\langle h^2\rangle/2$ and
$X_0\simeq \langle h\rangle$.   
Contrastingly, in a cracklike process, $\sqrt{X_1}$ and, hence, $\mathit{\Delta}X$
are proportional to $X_0$, because in such a process, a one-dimensional dry region develops. 
Figure \ref{DeltaX} displays a typical dependence of $\mathit{\Delta}X$ on
$X_0$.
As seen there, the invasion changes from that described by the
conventional IP model to cracklike invasion as $\tilde{G}$ increases. 
The exponent $\alpha'$, defined by $\mathit{\Delta}X\propto
X_0^{\alpha'}$ at $X_0\simeq 0$, is approximately $0.5$ for the invasion
of the conventional IP model and $1$ for cracklike invasion.

\begin{figure}
\includegraphics[width=7cm]{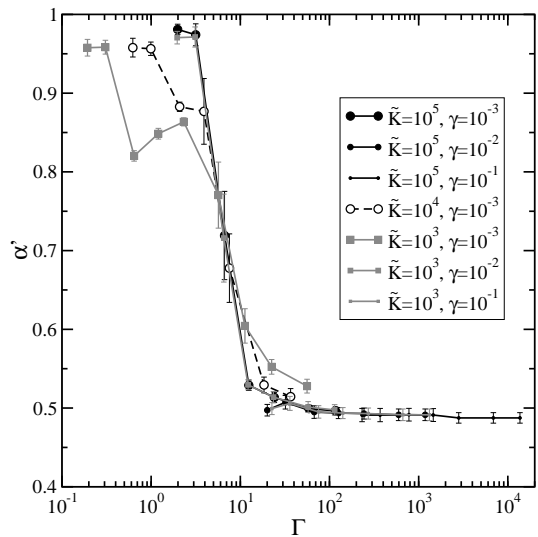}
 \caption{Dependence of $\alpha'$ on $\Gamma$. 
The data in Fig.~\ref{alpha-1/G} are replotted with respect to $\Gamma$, defined in
 Eq.~(\ref{Gamma}).}
\label{alpha-Gamma}
\end{figure}

\begin{figure}
\includegraphics[width=7.0cm]{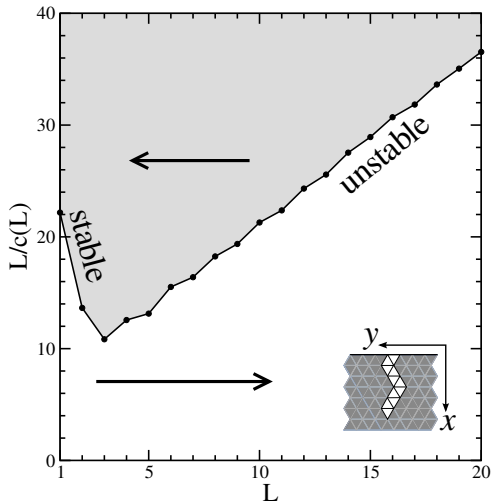} 
 \caption{$L/c(L)$ for a triangular lattice of $N_x=N_y=40$. 
The quantity $c(L)$ used here is the average over $100$ sets of $\{\gamma_m\}$ created from 
 different random seeds. The inset displays an example of a one-dimensional
 path with length $L=11$.}
\label{L_c} 
\end{figure}

Figure \ref{alpha-1/G} displays the exponent $\alpha'$ as a function of
$1/\tilde{G}$ for various $\tilde{K}$ and $\mathit{\Delta}\gamma$. 
The change of $\alpha'$
from $1.0$ to $0.5$ corresponds to the transition of the invasion type.  
The data are replotted with respect to $\Gamma$ in Fig.~\ref{alpha-Gamma}. 
It is seen that all data approximately collapse to a single curve for large $\tilde{K}$, and the transition 
occurs at approximately the same value of $\Gamma$ in each case, near $\Gamma \simeq 10$.
This result is consistent with the criterion appearing in
Eq.~(\ref{cracking condition}) and indicates that $L_c/c(L_c)\simeq
10$ in our simulations.

The function $c(L)$ is determined from the distribution function of $\gamma_m$ and the lattice properties.
Investigating all one-dimensional paths of length $L$, extending from the
top surface in the $x$ direction,  
we determined $c(L)$ from the minimum value of the sum of $\gamma_m$ over a path.
Figure \ref{L_c} displays $L/c(L)$ for the $N_x=N_y=40$ triangular lattice we used.
It is seen that, because $c(L)$ increases with $L$ from $\simeq 0$ at
$L=1$, $L/c(L)$ first decreases and then increases as $L$ increases from
$1$. 
For the portion of this graph in which $L/c(L)$ is increasing, the value
of $L$ along the curve is $L_c$, the crack length beyond which a crack
grows unstably. 
For the portion of this graph in which $L/c(L)$ is decreasing, the value
of $L$ along the curve corresponds to the depth of the dry region that develops following the first invasion due
       to heterogeneity. 
The minimum of $c(L)/L$ corresponds to the transition point of $\Gamma$ 
 below which $L_c$ vanishes and 
the first invasion spontaneously induces an initial crack that develops into cracklike invasion.
The value found here for $\Gamma$ is consistent with $\Gamma\simeq 10$,
obtained from Fig.~\ref{alpha-Gamma}.

The quantity $\Gamma$ generally depends on the particle size.
In the original, unscaled system, $(\tilde{K},\tilde{G})$ correspond 
to $(K,G)\equiv (p_{\gamma}/v_w^{\nu})(\tilde{K},\tilde{G})$.
For fixed elastic properties, represented by $(K,G)$ and 
heterogeneity, represented by $\mathit{\Delta}\gamma$,   
we find from Eq.~(\ref{p_gamma}) that $\Gamma$ given in 
Eq.~(\ref{Gamma}) with $\tilde{p}\simeq 1$ 
increases with the size of a particle as
 \begin{equation} 
\Gamma\simeq v_w p_{\gamma}^{-\frac{1}{\nu}} \frac{\mathit{\Delta}\gamma
 K^{\frac{1}{\nu}}}{g_{\nu}\left(\frac{G}{K}\right)}\propto
 r^{\frac{1}{\nu}}.
\label{Gamma:unscaled}
\end{equation}
This implies that cracking occurs only when the particle size is small.  
Similar results have been reported previously in some experiments \cite{Xu08}.

\subsection{Discussion on the validity}
We now discuss the validity of the cracking criterion given in
Eq.~(\ref{cracking condition}). 
This criterion seems valid for large $\tilde{K}$, even when shear bands
appear in advance of the first crack.
This is because this crack will simply follow the path of the shear
band. 
Most of the difference in the free energy is consumed when the shear
band forms, because a shear band enables 
compression in the surrounding region, due to the large strain it
creates.  
By contrast, cracks tend to become wide for sufficiently soft 
materials.
This is due to the fact that if a one-dimensional crack were to develop in such materials,  
the large expansion at a crack tip would cause additional dissipation, larger than $O((\mathit{\Delta}\gamma)^2)$, in
Eq.~(\ref{R:cracking}). 
The dependence of $\alpha'$ on $\Gamma$ seen in 
Fig.~\ref{alpha-Gamma} deviates from the master curve for small
$\tilde{K}$, because we assumed a crack to be a one-dimensional
region when we derived our criterion. 
In order to generalize our criterion to be applicable to such cases, we need to evaluate elastic strains
at the tip of a blunt crack. 

Cracking in pastelike materials is often accompanied by large
plastic relaxation~\cite{Goehring13} although we have ignored plasticity to elucidate the Griffith criterion
in this paper.  
Plastic deformation would increase additional dissipation significantly. 
If plastic deformation occurs locally in the vicinities of crack tips, our criterion could be 
extended to include dissipation energy in the same manner as in the 
standard fracture mechanics~\cite{Lawn93}. 
For pastelike materials, however, we should note that plastic deformation
 may occur globally by increasing negative pore pressures~\cite{Kitsunezaki10} and
 affect crack directions through the memory
 effects~\cite{Nakahara06,Nakahara11}. 

We have investigated the first crack appearing in a uniform layer without initial
cracks or flaws, except microscopic heterogeneity of drying properties. 
As $\Gamma$ does not depend on the layer 
thickness $H$ in Eq.~(\ref{Gamma:unscaled}), our results appear to
contradict the existence of the critical cracking thickness.    
However, if the layer contains an initial crack or a macroscopic air-invaded region initially, it can develop and divide
the system at smaller values of $\tilde{p}$. 
When $\tilde{p}$ is very small, the last term in Eq.~(\ref{Decrease of J for cracking}) can be approximated as 
$L\mathit{\Delta}V\{-\gamma_D(L)\}$, and the dissipation term can be 
ignored. 
In this case, the cracking condition can be written 
\begin{equation}
L\left\{\tilde{f}_e(\tilde{\bm{U}}_a)-\tilde{f}_e(\tilde{\bm{U}}_i)+\tilde{p}(\tilde{U}_a-\tilde{U}_i)\right\}
\gtrsim \gamma_D(\infty), 
\end{equation}
because $\gamma_D(L)$ is approximately constant for $L\gg 1$.
Assuming $L$ to be the layer thickness, $H$, 
this condition gives 
\begin{equation}
\tilde{p}\simeq \left(\frac{\tilde{K}^\frac{1}{\nu}\gamma_D(\infty)}{g_{\nu}\left(\frac{\tilde{G}}{\tilde{K}}\right)H}\right)^{\frac{\nu}{\nu+1}}
\end{equation}
as the smallest value of the pressure for which cracking can occur. 
For the original, unscaled system, we have $p \propto H^{-\frac{\nu}{\nu+1}}$, and 
$p$ does not depend on $r$, because $H$ is scaled by the unit of length,
which is proportional to $r$.
These dependencies are consistent with the theoretical and experimental
results obtained in previous research~\cite{Beuth92, Routh99, Man08, Singh07}.

\section{Conclusions}\label{Section:Conclusions}

We proposed an invasion percolation model for a cohesionless elastic 
material to investigate drying processes of pastelike materials. 
We derived a cracking condition that applies to cohesionless porous
systems taking the same form as the Griffith criterion, after 
eliminating local dissipation accompanied by air invasion.
The Griffith energy corresponds to an additional energy required for
drying, not the surface energy of the liquid-air interface itself.
We find that cracklike air invasion occurs for soft materials with larger
rigidity and less heterogeneity in the properties characterizing the
drying process.
Also, this criterion explains why cracking does not occur for systems
composed of large particles. 

For systems in which there is fast drying or large deformation, 
the cracking condition will differ from that derived here, because in 
such situations, there are complications that were not accounted for in
the present work. 
Specifically, in the case of fast drying, the pore pressure will become 
nonuniform, 
while in the case of larger deformation, plastic deformation will
appear. 
While it is important to elucidate such phenomena, these problems are
beyond the scope of the present work.

\begin{acknowledgments}
We thank H. Ito for her contribution to early numerical results.  
The author also acknowledges A.~Nakahara, Ooshida~T, T.~Mizuguchi,
 S.~Tarafder, T.~Dutta, and 
 C.~Urabe for useful discussions and G. C. Paquette for valuable
 comments.  
This research was supported by two Grants-in-Aid for Scientific
 Research (Grant No. KAKENHI C 23540452 and Grant No. KAKENHI B 22340112) from JSPS, Japan.
\end{acknowledgments}

\appendix
\section{Equilibrium conditions}\label{Equilibrium conditions}
Let us consider thermal equilibrium states of a paste system for 
given $(T,P,h)$, where the relative humidity, $h$, is given by
$h\equiv P_v/P_v^*$, where $P_v$ is the vapor pressure and $P_v^*$ is its saturated value. 
We assume that both the particles and the liquid composing the system are 
incompressible and that the vapor is an ideal gas, for simplicity.

A condition of mechanical equilibrium, Laplace's law, states
 that the liquid pressure, $P_l$, and the atmospheric pressure, $P$, are
 related as 
\begin{equation}
p\equiv P-P_l=\gamma_{la}\kappa. 
\end{equation}
Here $\gamma_{la}$ is the surface tension of the liquid-air interface and $\kappa$ is the mean curvature, which is constant everywhere on the
liquid-air interface in an equilibrium state. 

The negative pore pressure 
$p$ is determined from $h$ by the Kelvin condition,
\begin{equation}
p=-k_BT\rho_l \log{h},  
\label{Kelvin}
\end{equation}
for an ideal gas, where $k_B$ is Boltzmann's constant and 
$\rho_l\equiv N_l/V_l$ is the number density of liquid molecules.
This equation is derived from the chemical equilibrium condition
$\mu_l(T,P_l)=\mu_v(T,P_v)$ and the condition for a flat interface,
$\mu_l(T,P)=\mu_v(T,P_v^*)$, which provides the definition of $P_v^*$. 
In the derivation of Eq.~(\ref{Kelvin}), the relations $\partial \mu_l/\partial
P_l=1/\rho_l$ and $\mu_v=k_BT\log{P_v}+\mbox{const.}$ have been used.

Another mechanical equilibrium condition, Young-Dupr\'e's law,   
\begin{equation}
\gamma_{la}\cos{\theta}=\gamma_{sa}-\gamma_{sl}, 
 \label{Young-Dupre}
\end{equation}
holds at the contact points of the liquid-air interface and the
surfaces of solid particles.   
Here, $\theta$ is the contact angle. 
The interface energy of the paste is the sum of the surface energies of
the liquid-air, solid-air, and
solid-liquid interfaces. 
Explicitly, we have $F_i=\gamma _{la}A_{la}+\gamma_{sa}A_{sa}+\gamma_{sl}A_{sl}$,  
where $A_{mn}$ and $\gamma_{mn}$ are the surface area and the surface tension
of the interface indicated by their indices, respectively.
As the total area of the solid surfaces, $A_{sa}+A_{sl}$, is approximately
constant, $F_i$ can be written as  
\begin{equation}
F_i=\gamma_{la}A+\mbox{const.}, 
\label{F_i}
\end{equation}
where $A\equiv A_{la}+\cos{\theta}A_{sa}$,   
after substituting Eq.~(\ref{Young-Dupre}).
The invasion of air causes $A$ and thus $F_i$ to increase.

\section{Isotropic compressive states}\label{Isotropic compressive states}

The free energy $\tilde{J}$ of a wet cell is minimal with respect to
$\tilde{u}_{\alpha\beta}$ in the isotropic compressive state.
For Eqs.~(\ref{fe}) and (\ref{g:homogenerous}), $\tilde{J}$ depends on $\tilde{u}_{\alpha\beta}$ as 
\begin{equation}
\tilde{f}_e+\tilde{p}\tilde{u}_{\alpha\alpha}
=\frac{1}{2}(\tilde{K}-\tilde{G})(-\tilde{u}_{\alpha\alpha})^{\nu+1}+\tilde{G}(-\tilde{u}_{\eta\eta})^{\nu-1}\tilde{u}_{\alpha\beta}^2+\tilde{p}\tilde{u}_{\alpha\alpha}.
\end{equation}
When the strain tensor deviates from that of an isotropic state 
 by $U_{\alpha\beta}$, taking the form $\tilde{u}_{\alpha\beta}=\tilde{U}_i\left(\delta_{\alpha\beta}/2+U_{\alpha\beta}\right)$, this
 equation can be approximated to second order in $U_{\alpha\beta}$ as 
\begin{eqnarray}
\tilde{f}_e &+& \tilde{p}\tilde{u}_{\alpha\alpha}
\nonumber\\
&\simeq &\mbox{const.}
+\left\{\tilde{p}-\frac{\nu+1}{2}(-\tilde{U}_i)^{\nu}\tilde{K}\right\}\tilde{U}_iU_{\alpha\alpha}
\nonumber \\
&+&(-\tilde{U}_i)^{\nu+1}\left\{\frac{(\nu+1)\nu
 \tilde{K}-2\tilde{G}}{4}U_{\alpha\alpha}^2+\tilde{G}U_{\alpha\beta}^2\right\}.
\end{eqnarray}
Applying the condition that this quantity be minimal gives
Eq.~(\ref{Ui}) and $\tilde{K}, \tilde{G}>0$. 
The bulk modulus and rigidity in 2D linear elasticity 
are $\tilde{K}'\equiv (\nu+1)\nu(-\tilde{U}_i)^{\nu-1}\tilde{K}/2$ and
$\tilde{G}'\equiv (-\tilde{U}_i)^{\nu-1}\tilde{G}$, respectively, for the isotropic compressive state.  
Poisson's ratio is given by
$(\tilde{K}'-\tilde{G}')/(\tilde{K}'+\tilde{G}')$.

\section{Perturbation from a uniaxial compressive state}\label{Perturbation from a uniaxial compressive state}

A wet cell in contact with a dry-wet interface has the same stress
conditions, $\tilde{\sigma}_{xx}=\tilde{\sigma}_{xy}=0$, as 
the unixaial compressive state, described by $\tilde{\bm{U}}_a$ given in 
Eq.~(\ref{uniaxial compressive states}), where the $x$ axis is 
perpendicular to the interface, while  
the stress $\tilde{\sigma}_{yy}$ depends on the volumetric strain, $\tilde{u}_{\alpha\alpha}$.

The elastic energy is 
$\tilde{f}_e(\tilde{\bm{U}})=\tilde{f}_e(\tilde{\bm{U}}_a)
+\tilde{\sigma}_{e \alpha\beta}(\tilde{\bm{U}}_a)\delta
 \tilde{u}_{\alpha\beta}+O(\delta \tilde{\bm{U}}^2)$ for 
 $\tilde{\bm{U}}\equiv{\tilde{\bm{U}}_a}+\delta\tilde{\bm{U}}$, where 
 $\tilde{\sigma}_{e \alpha\beta}\equiv \partial \tilde{f}_e/\partial
\tilde{u}_{\alpha\beta}=\tilde{\sigma}_{\alpha\beta}-\tilde{p}\delta_{\alpha\beta}$.
The first-order term can be rewritten as  
\begin{eqnarray}
\tilde{\sigma}_{e \alpha\beta}\delta 
\tilde{u}_{\alpha\beta} 
=&& \frac{1}{\nu}\tilde{u}_{\alpha\beta}\delta
 \tilde{\sigma}_{e \alpha\beta}
\nonumber \\
=&& \frac{1}{\nu}
(\tilde{u}_{xx}\delta\tilde{\sigma}_{e
xx}+\tilde{u}_{xy}\delta\tilde{\sigma}_{e xy}+\tilde{u}_{yy}\delta
\tilde{\sigma}_{e yy}),  
\label{sigma u}
\end{eqnarray}
because 
$\tilde{f}_e=\tilde{\sigma}_{e\alpha\beta}\tilde{u}_{\alpha\beta}/(\nu+1)$ for the homogeneous
function given in Eq.~(\ref{g:homogenerous}) and  
$d\tilde{f}_e=\tilde{\sigma}_{e \alpha\beta}d
\tilde{u}_{\alpha\beta}=(\tilde{\sigma}_{e \alpha\beta}d\tilde{u}_{\alpha\beta}+d
\tilde{\sigma}_{e \alpha\beta}\tilde{u}_{\alpha\beta})/(1+\nu)$. 
Equation (\ref{sigma u}) vanishes, because $\delta \tilde{\sigma}_{e xx}=\delta \tilde{\sigma}_{e xy}=0$
 for fixed $\tilde{p}$, due to the stress conditions and  $\tilde{u}_{yy}(\tilde{\bm{U}}_a)=0$.
Thus, we have  
$\tilde{f}_e(\tilde{\bm{U}})-\tilde{f}_e(\tilde{\bm{U}}_a)
=O(\delta \tilde{\bm{U}}^2)$ 
and 
\begin{equation}
\tilde{\sigma}_{e \alpha\beta}\delta
 \tilde{u}_{\alpha\beta}=-\tilde{p}\delta\tilde{u}_{xx}+\tilde{\sigma}_{e
 yy}\delta
\tilde{u}_{yy}=0. 
\label{simga u=0}
\end{equation}
The quantity $\delta\tilde{\bm{U}}$ is determined from Eq.~(\ref{simga u=0}) and $\delta\tilde{\sigma}_{xy}=0$ 
 for a given
 $\delta\tilde{u}_{\alpha\alpha}=\delta\tilde{u}_{xx}+\delta\tilde{u}_{yy}$, 
and thus we find 
\begin{equation}
\tilde{f}_e(\tilde{\bm{U}})-\tilde{f}_e({\bm
 U}_a)=O((\delta\tilde{u}_{\alpha\alpha})^2).
\end{equation}


\begin{thebibliography}{44}%
\makeatletter
\providecommand \@ifxundefined [1]{%
 \@ifx{#1\undefined}
}%
\providecommand \@ifnum [1]{%
 \ifnum #1\expandafter \@firstoftwo
 \else \expandafter \@secondoftwo
 \fi
}%
\providecommand \@ifx [1]{%
 \ifx #1\expandafter \@firstoftwo
 \else \expandafter \@secondoftwo
 \fi
}%
\providecommand \natexlab [1]{#1}%
\providecommand \enquote  [1]{``#1''}%
\providecommand \bibnamefont  [1]{#1}%
\providecommand \bibfnamefont [1]{#1}%
\providecommand \citenamefont [1]{#1}%
\providecommand \href@noop [0]{\@secondoftwo}%
\providecommand \href [0]{\begingroup \@sanitize@url \@href}%
\providecommand \@href[1]{\@@startlink{#1}\@@href}%
\providecommand \@@href[1]{\endgroup#1\@@endlink}%
\providecommand \@sanitize@url [0]{\catcode `\\12\catcode `\$12\catcode
  `\&12\catcode `\#12\catcode `\^12\catcode `\_12\catcode `\%12\relax}%
\providecommand \@@startlink[1]{}%
\providecommand \@@endlink[0]{}%
\providecommand \url  [0]{\begingroup\@sanitize@url \@url }%
\providecommand \@url [1]{\endgroup\@href {#1}{\urlprefix }}%
\providecommand \urlprefix  [0]{URL }%
\providecommand \Eprint [0]{\href }%
\providecommand \doibase [0]{http://dx.doi.org/}%
\providecommand \selectlanguage [0]{\@gobble}%
\providecommand \bibinfo  [0]{\@secondoftwo}%
\providecommand \bibfield  [0]{\@secondoftwo}%
\providecommand \translation [1]{[#1]}%
\providecommand \BibitemOpen [0]{}%
\providecommand \bibitemStop [0]{}%
\providecommand \bibitemNoStop [0]{.\EOS\space}%
\providecommand \EOS [0]{\spacefactor3000\relax}%
\providecommand \BibitemShut  [1]{\csname bibitem#1\endcsname}%
\let\auto@bib@innerbib\@empty
\bibitem [{\citenamefont {Pauchard}\ \emph {et~al.}(2003)\citenamefont
  {Pauchard}, \citenamefont {Adda-Bedia}, \citenamefont {Allain},\ and\
  \citenamefont {Couder}}]{Pauchard03}%
  \BibitemOpen
  \bibfield  {author} {\bibinfo {author} {\bibfnamefont {L.}~\bibnamefont
  {Pauchard}}, \bibinfo {author} {\bibfnamefont {M.}~\bibnamefont
  {Adda-Bedia}}, \bibinfo {author} {\bibfnamefont {C.}~\bibnamefont {Allain}},
  \ and\ \bibinfo {author} {\bibfnamefont {Y.}~\bibnamefont {Couder}},\
  }\href@noop {} {\bibfield  {journal} {\bibinfo  {journal} {Phys. Rev. E}\
  }\textbf {\bibinfo {volume} {67}},\ \bibinfo {pages} {027103} (\bibinfo
  {year} {2003})}\BibitemShut {NoStop}%
\bibitem [{\citenamefont {Bohn}\ \emph
  {et~al.}(2005{\natexlab{a}})\citenamefont {Bohn}, \citenamefont {Pauchard},\
  and\ \citenamefont {Couder}}]{Bohn05}%
  \BibitemOpen
  \bibfield  {author} {\bibinfo {author} {\bibfnamefont {S.}~\bibnamefont
  {Bohn}}, \bibinfo {author} {\bibfnamefont {L.}~\bibnamefont {Pauchard}}, \
  and\ \bibinfo {author} {\bibfnamefont {Y.}~\bibnamefont {Couder}},\
  }\href@noop {} {\bibfield  {journal} {\bibinfo  {journal} {Phys. Rev. E}\
  }\textbf {\bibinfo {volume} {71}},\ \bibinfo {pages} {046214} (\bibinfo
  {year} {2005}{\natexlab{a}})}\BibitemShut {NoStop}%
\bibitem [{\citenamefont {Bohn}\ \emph
  {et~al.}(2005{\natexlab{b}})\citenamefont {Bohn}, \citenamefont
  {Platkiewicz}, \citenamefont {Andreotti}, \citenamefont {Adda-Bedia},\ and\
  \citenamefont {Couder}}]{Bohn05b}%
  \BibitemOpen
  \bibfield  {author} {\bibinfo {author} {\bibfnamefont {S.}~\bibnamefont
  {Bohn}}, \bibinfo {author} {\bibfnamefont {J.}~\bibnamefont {Platkiewicz}},
  \bibinfo {author} {\bibfnamefont {B.}~\bibnamefont {Andreotti}}, \bibinfo
  {author} {\bibfnamefont {M.}~\bibnamefont {Adda-Bedia}}, \ and\ \bibinfo
  {author} {\bibfnamefont {Y.}~\bibnamefont {Couder}},\ }\href@noop {}
  {\bibfield  {journal} {\bibinfo  {journal} {Phys. Rev. E}\ }\textbf {\bibinfo
  {volume} {71}},\ \bibinfo {pages} {046215} (\bibinfo {year}
  {2005}{\natexlab{b}})}\BibitemShut {NoStop}%
\bibitem [{\citenamefont {Nakahara}\ and\ \citenamefont
  {Matsuo}(2006)}]{Nakahara06}%
  \BibitemOpen
  \bibfield  {author} {\bibinfo {author} {\bibfnamefont {A.}~\bibnamefont
  {Nakahara}}\ and\ \bibinfo {author} {\bibfnamefont {Y.}~\bibnamefont
  {Matsuo}},\ }\href@noop {} {\bibfield  {journal} {\bibinfo  {journal} {Phys.
  Rev. E}\ }\textbf {\bibinfo {volume} {74}},\ \bibinfo {pages} {045102(R)}
  (\bibinfo {year} {2006})}\BibitemShut {NoStop}%
\bibitem [{\citenamefont {Mal}\ \emph {et~al.}(2008)\citenamefont {Mal},
  \citenamefont {Sinha}, \citenamefont {Middya},\ and\ \citenamefont
  {Tarafdar}}]{Mal08}%
  \BibitemOpen
  \bibfield  {author} {\bibinfo {author} {\bibfnamefont {D.}~\bibnamefont
  {Mal}}, \bibinfo {author} {\bibfnamefont {S.}~\bibnamefont {Sinha}}, \bibinfo
  {author} {\bibfnamefont {T.~R.}\ \bibnamefont {Middya}}, \ and\ \bibinfo
  {author} {\bibfnamefont {S.}~\bibnamefont {Tarafdar}},\ }\href@noop {}
  {\bibfield  {journal} {\bibinfo  {journal} {Appl. Clay Sci.}\ }\textbf
  {\bibinfo {volume} {39}},\ \bibinfo {pages} {106} (\bibinfo {year}
  {2008})}\BibitemShut {NoStop}%
\bibitem [{\citenamefont {Pauchard}\ \emph {et~al.}(2009)\citenamefont
  {Pauchard}, \citenamefont {Abou},\ and\ \citenamefont
  {Sekimoto}}]{Pauchard09}%
  \BibitemOpen
  \bibfield  {author} {\bibinfo {author} {\bibfnamefont {L.}~\bibnamefont
  {Pauchard}}, \bibinfo {author} {\bibfnamefont {B.}~\bibnamefont {Abou}}, \
  and\ \bibinfo {author} {\bibfnamefont {K.}~\bibnamefont {Sekimoto}},\
  }\href@noop {} {\bibfield  {journal} {\bibinfo  {journal} {Langmuir}\
  }\textbf {\bibinfo {volume} {25}},\ \bibinfo {pages} {6672} (\bibinfo {year}
  {2009})}\BibitemShut {NoStop}%
\bibitem [{\citenamefont {Kitsunezaki}(2009)}]{Kitsunezaki09}%
  \BibitemOpen
  \bibfield  {author} {\bibinfo {author} {\bibfnamefont {S.}~\bibnamefont
  {Kitsunezaki}},\ }\href@noop {} {\bibfield  {journal} {\bibinfo  {journal}
  {J. Phys. Soc. Jpn.}\ }\textbf {\bibinfo {volume} {78}},\ \bibinfo {pages}
  {064801} (\bibinfo {year} {2009})}\BibitemShut {NoStop}%
\bibitem [{\citenamefont {Kitsunezaki}(2010)}]{Kitsunezaki10}%
  \BibitemOpen
  \bibfield  {author} {\bibinfo {author} {\bibfnamefont {S.}~\bibnamefont
  {Kitsunezaki}},\ }\href@noop {} {\bibfield  {journal} {\bibinfo  {journal}
  {J. Phys. Soc. Jpn.}\ }\textbf {\bibinfo {volume} {79}},\ \bibinfo {pages}
  {124802} (\bibinfo {year} {2010})}\BibitemShut {NoStop}%
\bibitem [{\citenamefont {Goehring}\ \emph
  {et~al.}(2010{\natexlab{a}})\citenamefont {Goehring}, \citenamefont {Conroy},
  \citenamefont {Akhter}, \citenamefont {Clegg},\ and\ \citenamefont
  {Routh}}]{Goehring10a}%
  \BibitemOpen
  \bibfield  {author} {\bibinfo {author} {\bibfnamefont {L.}~\bibnamefont
  {Goehring}}, \bibinfo {author} {\bibfnamefont {R.}~\bibnamefont {Conroy}},
  \bibinfo {author} {\bibfnamefont {A.}~\bibnamefont {Akhter}}, \bibinfo
  {author} {\bibfnamefont {W.~J.}\ \bibnamefont {Clegg}}, \ and\ \bibinfo
  {author} {\bibfnamefont {A.~F.}\ \bibnamefont {Routh}},\ }\href@noop {}
  {\bibfield  {journal} {\bibinfo  {journal} {Soft Matter}\ }\textbf {\bibinfo
  {volume} {6}},\ \bibinfo {pages} {3562} (\bibinfo {year}
  {2010}{\natexlab{a}})}\BibitemShut {NoStop}%
\bibitem [{\citenamefont {Goehring}\ \emph
  {et~al.}(2010{\natexlab{b}})\citenamefont {Goehring}, \citenamefont {Clegg},\
  and\ \citenamefont {Routh}}]{Goehring10b}%
  \BibitemOpen
  \bibfield  {author} {\bibinfo {author} {\bibfnamefont {L.}~\bibnamefont
  {Goehring}}, \bibinfo {author} {\bibfnamefont {W.~J.}\ \bibnamefont {Clegg}},
  \ and\ \bibinfo {author} {\bibfnamefont {A.~F.}\ \bibnamefont {Routh}},\
  }\href@noop {} {\bibfield  {journal} {\bibinfo  {journal} {Langmuir}\
  }\textbf {\bibinfo {volume} {26}},\ \bibinfo {pages} {9269} (\bibinfo {year}
  {2010}{\natexlab{b}})}\BibitemShut {NoStop}%
\bibitem [{\citenamefont {Goehring}\ \emph {et~al.}(2013)\citenamefont
  {Goehring}, \citenamefont {Clegg},\ and\ \citenamefont {Routh}}]{Goehring13}%
  \BibitemOpen
  \bibfield  {author} {\bibinfo {author} {\bibfnamefont {L.}~\bibnamefont
  {Goehring}}, \bibinfo {author} {\bibfnamefont {W.~J.}\ \bibnamefont {Clegg}},
  \ and\ \bibinfo {author} {\bibfnamefont {A.~F.}\ \bibnamefont {Routh}},\
  }\href@noop {} {\bibfield  {journal} {\bibinfo  {journal} {Phys. Rev. Lett.}\
  }\textbf {\bibinfo {volume} {110}},\ \bibinfo {pages} {024301} (\bibinfo
  {year} {2013})}\BibitemShut {NoStop}%
\bibitem [{\citenamefont {Nakahara}\ \emph {et~al.}(2011)\citenamefont
  {Nakahara}, \citenamefont {Shinohara},\ and\ \citenamefont
  {Matsuo}}]{Nakahara11}%
  \BibitemOpen
  \bibfield  {author} {\bibinfo {author} {\bibfnamefont {A.}~\bibnamefont
  {Nakahara}}, \bibinfo {author} {\bibfnamefont {Y.}~\bibnamefont {Shinohara}},
  \ and\ \bibinfo {author} {\bibfnamefont {Y.}~\bibnamefont {Matsuo}},\
  }\href@noop {} {\bibfield  {journal} {\bibinfo  {journal} {J. Phys. : Conf.
  Ser.}\ }\textbf {\bibinfo {volume} {319}},\ \bibinfo {pages} {012014}
  (\bibinfo {year} {2011})}\BibitemShut {NoStop}%
\bibitem [{\citenamefont {Routh}\ and\ \citenamefont {Russel}(1999)}]{Routh99}%
  \BibitemOpen
  \bibfield  {author} {\bibinfo {author} {\bibfnamefont {A.~F.}\ \bibnamefont
  {Routh}}\ and\ \bibinfo {author} {\bibfnamefont {W.~B.}\ \bibnamefont
  {Russel}},\ }\href@noop {} {\bibfield  {journal} {\bibinfo  {journal}
  {Langmuir}\ }\textbf {\bibinfo {volume} {15}},\ \bibinfo {pages} {7762}
  (\bibinfo {year} {1999})}\BibitemShut {NoStop}%
\bibitem [{\citenamefont {Tirumkudulu}\ and\ \citenamefont
  {Russel}(2005)}]{Tirumkudulu05}%
  \BibitemOpen
  \bibfield  {author} {\bibinfo {author} {\bibfnamefont {M.~S.}\ \bibnamefont
  {Tirumkudulu}}\ and\ \bibinfo {author} {\bibfnamefont {W.~B.}\ \bibnamefont
  {Russel}},\ }\href@noop {} {\bibfield  {journal} {\bibinfo  {journal}
  {Langmuir}\ }\textbf {\bibinfo {volume} {21}},\ \bibinfo {pages} {4938}
  (\bibinfo {year} {2005})}\BibitemShut {NoStop}%
\bibitem [{\citenamefont {Singh}\ and\ \citenamefont
  {Tirumkudulu}(2007)}]{Singh07}%
  \BibitemOpen
  \bibfield  {author} {\bibinfo {author} {\bibfnamefont {K.~B.}\ \bibnamefont
  {Singh}}\ and\ \bibinfo {author} {\bibfnamefont {M.~S.}\ \bibnamefont
  {Tirumkudulu}},\ }\href@noop {} {\bibfield  {journal} {\bibinfo  {journal}
  {Phys. Rev. Lett.}\ }\textbf {\bibinfo {volume} {98}},\ \bibinfo {pages}
  {218302} (\bibinfo {year} {2007})}\BibitemShut {NoStop}%
\bibitem [{\citenamefont {Russel}\ \emph {et~al.}(2008)\citenamefont {Russel},
  \citenamefont {Wu},\ and\ \citenamefont {Man}}]{Russel08}%
  \BibitemOpen
  \bibfield  {author} {\bibinfo {author} {\bibfnamefont {W.~B.}\ \bibnamefont
  {Russel}}, \bibinfo {author} {\bibfnamefont {N.}~\bibnamefont {Wu}}, \ and\
  \bibinfo {author} {\bibfnamefont {W.}~\bibnamefont {Man}},\ }\href@noop {}
  {\bibfield  {journal} {\bibinfo  {journal} {Langmuir}\ }\textbf {\bibinfo
  {volume} {24}},\ \bibinfo {pages} {1721} (\bibinfo {year}
  {2008})}\BibitemShut {NoStop}%
\bibitem [{\citenamefont {Man}\ and\ \citenamefont {Russel}(2008)}]{Man08}%
  \BibitemOpen
  \bibfield  {author} {\bibinfo {author} {\bibfnamefont {W.}~\bibnamefont
  {Man}}\ and\ \bibinfo {author} {\bibfnamefont {W.~B.}\ \bibnamefont
  {Russel}},\ }\href@noop {} {\bibfield  {journal} {\bibinfo  {journal} {Phys.
  Rev. Lett.}\ }\textbf {\bibinfo {volume} {100}},\ \bibinfo {pages} {198302}
  (\bibinfo {year} {2008})}\BibitemShut {NoStop}%
\bibitem [{\citenamefont {Wood}(1990)}]{Wood90}%
  \BibitemOpen
  \bibfield  {author} {\bibinfo {author} {\bibfnamefont {D.~M.}\ \bibnamefont
  {Wood}},\ }\href@noop {} {\emph {\bibinfo {title} {Soil Behaviour and
  Critical State Soil Mechanics}}}\ (\bibinfo  {publisher} {Cambridge
  University Press},\ \bibinfo {address} {New York},\ \bibinfo {year}
  {1990})\BibitemShut {NoStop}%
\bibitem [{\citenamefont {Otsuki}(2005)}]{Otsuki05}%
  \BibitemOpen
  \bibfield  {author} {\bibinfo {author} {\bibfnamefont {M.}~\bibnamefont
  {Otsuki}},\ }\href@noop {} {\bibfield  {journal} {\bibinfo  {journal} {Phys.
  Rev. E}\ }\textbf {\bibinfo {volume} {72}},\ \bibinfo {pages} {046115}
  (\bibinfo {year} {2005})}\BibitemShut {NoStop}%
\bibitem [{\citenamefont {Takeshi}(2008)}]{Ooshida09}%
  \BibitemOpen
  \bibfield  {author} {\bibinfo {author} {\bibfnamefont {Ooshida}~\bibnamefont
  {Takeshi}},\ }\href@noop {} {\bibfield  {journal} {\bibinfo  {journal} {Phys.
  Rev. E}\ }\textbf {\bibinfo {volume} {77}},\ \bibinfo {pages} {061501}
  (\bibinfo {year} {2008})}\BibitemShut {NoStop}%
\bibitem [{\citenamefont {Wilkinson}\ and\ \citenamefont
  {Willemsen}(1983)}]{Wilkinson83}%
  \BibitemOpen
  \bibfield  {author} {\bibinfo {author} {\bibfnamefont {D.}~\bibnamefont
  {Wilkinson}}\ and\ \bibinfo {author} {\bibfnamefont {J.~F.}\ \bibnamefont
  {Willemsen}},\ }\href@noop {} {\bibfield  {journal} {\bibinfo  {journal} {J.
  Phys. A}\ }\textbf {\bibinfo {volume} {16}},\ \bibinfo {pages} {3365}
  (\bibinfo {year} {1983})}\BibitemShut {NoStop}%
\bibitem [{\citenamefont {Wilkinson}(1984)}]{Wilkinson84}%
  \BibitemOpen
  \bibfield  {author} {\bibinfo {author} {\bibfnamefont {D.}~\bibnamefont
  {Wilkinson}},\ }\href@noop {} {\bibfield  {journal} {\bibinfo  {journal}
  {Phys. Rev. A}\ }\textbf {\bibinfo {volume} {30}},\ \bibinfo {pages} {520}
  (\bibinfo {year} {1984})}\BibitemShut {NoStop}%
\bibitem [{\citenamefont {Wilkinson}(1986)}]{Wilkinson86}%
  \BibitemOpen
  \bibfield  {author} {\bibinfo {author} {\bibfnamefont {D.}~\bibnamefont
  {Wilkinson}},\ }\href@noop {} {\bibfield  {journal} {\bibinfo  {journal}
  {Phys. Rev. A}\ }\textbf {\bibinfo {volume} {34}},\ \bibinfo {pages} {1380}
  (\bibinfo {year} {1986})}\BibitemShut {NoStop}%
\bibitem [{\citenamefont {Du}\ \emph {et~al.}(1995)\citenamefont {Du},
  \citenamefont {Xu}, \citenamefont {Yortsos}, \citenamefont {Chaouche},
  \citenamefont {Rakotomalala},\ and\ \citenamefont {Salin}}]{Du95}%
  \BibitemOpen
  \bibfield  {author} {\bibinfo {author} {\bibfnamefont {C.}~\bibnamefont
  {Du}}, \bibinfo {author} {\bibfnamefont {B.}~\bibnamefont {Xu}}, \bibinfo
  {author} {\bibfnamefont {Y.~C.}\ \bibnamefont {Yortsos}}, \bibinfo {author}
  {\bibfnamefont {M.}~\bibnamefont {Chaouche}}, \bibinfo {author}
  {\bibfnamefont {N.}~\bibnamefont {Rakotomalala}}, \ and\ \bibinfo {author}
  {\bibfnamefont {D.}~\bibnamefont {Salin}},\ }\href@noop {} {\bibfield
  {journal} {\bibinfo  {journal} {Phys. Rev. Lett.}\ }\textbf {\bibinfo
  {volume} {74}},\ \bibinfo {pages} {694} (\bibinfo {year} {1995})}\BibitemShut
  {NoStop}%
\bibitem [{\citenamefont {Yortsos}\ \emph {et~al.}(1997)\citenamefont
  {Yortsos}, \citenamefont {Xu},\ and\ \citenamefont {Salin}}]{Yortsos97}%
  \BibitemOpen
  \bibfield  {author} {\bibinfo {author} {\bibfnamefont {Y.~C.}\ \bibnamefont
  {Yortsos}}, \bibinfo {author} {\bibfnamefont {B.}~\bibnamefont {Xu}}, \ and\
  \bibinfo {author} {\bibfnamefont {D.}~\bibnamefont {Salin}},\ }\href@noop {}
  {\bibfield  {journal} {\bibinfo  {journal} {Phys. Rev. Lett.}\ }\textbf
  {\bibinfo {volume} {79}},\ \bibinfo {pages} {4581} (\bibinfo {year}
  {1997})}\BibitemShut {NoStop}%
\bibitem [{\citenamefont {Meakin}\ \emph {et~al.}(2000)\citenamefont {Meakin},
  \citenamefont {Wagner}, \citenamefont {Vedvik}, \citenamefont {Amundsen},
  \citenamefont {Feder},\ and\ \citenamefont {J{\o}ssang}}]{Meakin00}%
  \BibitemOpen
  \bibfield  {author} {\bibinfo {author} {\bibfnamefont {P.}~\bibnamefont
  {Meakin}}, \bibinfo {author} {\bibfnamefont {G.}~\bibnamefont {Wagner}},
  \bibinfo {author} {\bibfnamefont {A.}~\bibnamefont {Vedvik}}, \bibinfo
  {author} {\bibfnamefont {H.}~\bibnamefont {Amundsen}}, \bibinfo {author}
  {\bibfnamefont {J.}~\bibnamefont {Feder}}, \ and\ \bibinfo {author}
  {\bibfnamefont {T.}~\bibnamefont {J{\o}ssang}},\ }\href@noop {} {\bibfield
  {journal} {\bibinfo  {journal} {Mar. Petrol. Geol.}\ }\textbf {\bibinfo
  {volume} {17}},\ \bibinfo {pages} {777} (\bibinfo {year} {2000})}\BibitemShut
  {NoStop}%
\bibitem [{\citenamefont {L{\o}voll}\ \emph {et~al.}(2005)\citenamefont
  {L{\o}voll}, \citenamefont {M\'{e}heust}, \citenamefont {M{\aa}l{\o}ya},
  \citenamefont {Aker},\ and\ \citenamefont {Schmittbuhl}}]{Lovoll05}%
  \BibitemOpen
  \bibfield  {author} {\bibinfo {author} {\bibfnamefont {G.}~\bibnamefont
  {L{\o}voll}}, \bibinfo {author} {\bibfnamefont {Y.}~\bibnamefont
  {M\'{e}heust}}, \bibinfo {author} {\bibfnamefont {K.~J.}\ \bibnamefont
  {M{\aa}l{\o}ya}}, \bibinfo {author} {\bibfnamefont {E.}~\bibnamefont {Aker}},
  \ and\ \bibinfo {author} {\bibfnamefont {J.}~\bibnamefont {Schmittbuhl}},\
  }\href@noop {} {\bibfield  {journal} {\bibinfo  {journal} {Energy}\ }\textbf
  {\bibinfo {volume} {30}},\ \bibinfo {pages} {861} (\bibinfo {year}
  {2005})}\BibitemShut {NoStop}%
\bibitem [{\citenamefont {Yamazaki}\ \emph {et~al.}(2006)\citenamefont
  {Yamazaki}, \citenamefont {Komura},\ and\ \citenamefont
  {Suganuma}}]{Yamazaki06}%
  \BibitemOpen
  \bibfield  {author} {\bibinfo {author} {\bibfnamefont {Y.}~\bibnamefont
  {Yamazaki}}, \bibinfo {author} {\bibfnamefont {S.}~\bibnamefont {Komura}}, \
  and\ \bibinfo {author} {\bibfnamefont {K.}~\bibnamefont {Suganuma}},\
  }\href@noop {} {\bibfield  {journal} {\bibinfo  {journal} {J. Phys. Soc.
  Jpn.}\ }\textbf {\bibinfo {volume} {75}},\ \bibinfo {pages} {043001}
  (\bibinfo {year} {2006})}\BibitemShut {NoStop}%
\bibitem [{\citenamefont {Nakanishi}\ \emph {et~al.}(2007)\citenamefont
  {Nakanishi}, \citenamefont {Yamamoto}, \citenamefont {Hayase},\ and\
  \citenamefont {Mitarai}}]{Nakanishi07}%
  \BibitemOpen
  \bibfield  {author} {\bibinfo {author} {\bibfnamefont {H.}~\bibnamefont
  {Nakanishi}}, \bibinfo {author} {\bibfnamefont {R.}~\bibnamefont {Yamamoto}},
  \bibinfo {author} {\bibfnamefont {Y.}~\bibnamefont {Hayase}}, \ and\ \bibinfo
  {author} {\bibfnamefont {N.}~\bibnamefont {Mitarai}},\ }\href@noop {}
  {\bibfield  {journal} {\bibinfo  {journal} {J. Phys. Soc. Jpn.}\ }\textbf
  {\bibinfo {volume} {76}},\ \bibinfo {pages} {024003} (\bibinfo {year}
  {2007})}\BibitemShut {NoStop}%
\bibitem [{\citenamefont {Xu}\ \emph {et~al.}(2008)\citenamefont {Xu},
  \citenamefont {Davies}, \citenamefont {Schofield},\ and\ \citenamefont
  {Weitz}}]{Xu08}%
  \BibitemOpen
  \bibfield  {author} {\bibinfo {author} {\bibfnamefont {L.}~\bibnamefont
  {Xu}}, \bibinfo {author} {\bibfnamefont {S.}~\bibnamefont {Davies}}, \bibinfo
  {author} {\bibfnamefont {A.~B.}\ \bibnamefont {Schofield}}, \ and\ \bibinfo
  {author} {\bibfnamefont {D.~A.}\ \bibnamefont {Weitz}},\ }\href@noop {}
  {\bibfield  {journal} {\bibinfo  {journal} {Phys. Rev. Lett.}\ }\textbf
  {\bibinfo {volume} {101}},\ \bibinfo {pages} {094502} (\bibinfo {year}
  {2008})}\BibitemShut {NoStop}%
\bibitem [{\citenamefont {Shokri}\ and\ \citenamefont {Or}(2013)}]{Shokri13}%
  \BibitemOpen
  \bibfield  {author} {\bibinfo {author} {\bibfnamefont {N.}~\bibnamefont
  {Shokri}}\ and\ \bibinfo {author} {\bibfnamefont {D.}~\bibnamefont {Or}},\
  }\href@noop {} {\bibfield  {journal} {\bibinfo  {journal} {J. Colloid
  Interface Sci.}\ }\textbf {\bibinfo {volume} {391}},\ \bibinfo {pages}
  {135} (\bibinfo {year} {2013})}\BibitemShut {NoStop}%
\bibitem [{Note1()}]{Note1}%
  \BibitemOpen
  \bibinfo {note} {Strong cohesion is often formed among constituent particles,
  typically after drying~\cite {Kendall87}. In such cases, cracking has common
  properties with typical brittle fracture in contrast to that in capillary
  states~\cite {Kitsunezaki09}.}\BibitemShut {Stop}%
\bibitem [{\citenamefont {Goehring}(2009)}]{Goehring09}%
  \BibitemOpen
  \bibfield  {author} {\bibinfo {author} {\bibfnamefont {L.}~\bibnamefont
  {Goehring}},\ }\href@noop {} {\bibfield  {journal} {\bibinfo  {journal}
  {Phys. Rev. E}\ }\textbf {\bibinfo {volume} {80}},\ \bibinfo {pages} {036116}
  (\bibinfo {year} {2009})}\BibitemShut {NoStop}%
\bibitem [{\citenamefont {Jiang}\ and\ \citenamefont {Liu}(2007)}]{Jiang07}%
  \BibitemOpen
  \bibfield  {author} {\bibinfo {author} {\bibfnamefont {Y.}~\bibnamefont
  {Jiang}}\ and\ \bibinfo {author} {\bibfnamefont {M.}~\bibnamefont {Liu}},\
  }\href@noop {} {\bibfield  {journal} {\bibinfo  {journal} {Eur. Phys. J. E}\
  }\textbf {\bibinfo {volume} {22}},\ \bibinfo {pages} {255} (\bibinfo {year}
  {2007})}\BibitemShut {NoStop}%
\bibitem [{\citenamefont {Kitsunezaki}(1999)}]{Kitsunezaki99}%
  \BibitemOpen
  \bibfield  {author} {\bibinfo {author} {\bibfnamefont {S.}~\bibnamefont
  {Kitsunezaki}},\ }\href@noop {} {\bibfield  {journal} {\bibinfo  {journal}
  {Phys. Rev. E}\ }\textbf {\bibinfo {volume} {60}},\ \bibinfo {pages} {6449}
  (\bibinfo {year} {1999})}\BibitemShut {NoStop}%
\bibitem [{\citenamefont {Press}\ \emph {et~al.}(2002)\citenamefont {Press},
  \citenamefont {Teukolsky}, \citenamefont {Vetterling},\ and\ \citenamefont
  {Flannery}}]{Press02}%
  \BibitemOpen
  \bibfield  {author} {\bibinfo {author} {\bibfnamefont {W.~H.}\ \bibnamefont
  {Press}}, \bibinfo {author} {\bibfnamefont {S.~A.}\ \bibnamefont
  {Teukolsky}}, \bibinfo {author} {\bibfnamefont {W.~T.}\ \bibnamefont
  {Vetterling}}, \ and\ \bibinfo {author} {\bibfnamefont {B.~P.}\ \bibnamefont
  {Flannery}},\ }\href@noop {} {\emph {\bibinfo {title} {Numerical Recipes in
  C++, 2nd ed.}}}\ (\bibinfo  {publisher} {Cambridge University Press},\ \bibinfo
  {address} {New York},\ \bibinfo {year} {2002})\BibitemShut {NoStop}%
\bibitem [{\citenamefont {Chakrabarti}\ and\ \citenamefont
  {Benguigui}(1997)}]{Chakrabarti97}%
  \BibitemOpen
  \bibfield  {author} {\bibinfo {author} {\bibfnamefont {B.~K.}\ \bibnamefont
  {Chakrabarti}}\ and\ \bibinfo {author} {\bibfnamefont {L.~G.}\ \bibnamefont
  {Benguigui}},\ }\href@noop {} {\emph {\bibinfo {title} {Statistical Physics
  of Fracture and Breakdown in Disordered Systems}}}\ (\bibinfo  {publisher}
  {Oxford University Press},\ \bibinfo {address} {New York},\ \bibinfo {year}
  {1997})\BibitemShut {NoStop}%
\bibitem [{\citenamefont {Shekhawat}\ \emph {et~al.}(2012)\citenamefont
  {Shekhawat}, \citenamefont {Zapperi},\ and\ \citenamefont
  {Sethna}}]{Shekhawat12}%
  \BibitemOpen
  \bibfield  {author} {\bibinfo {author} {\bibfnamefont {A.}~\bibnamefont
  {Shekhawat}}, \bibinfo {author} {\bibfnamefont {S.}~\bibnamefont {Zapperi}},
  \ and\ \bibinfo {author} {\bibfnamefont {J.~P.}\ \bibnamefont {Sethna}},\
  }\href@noop {} 
  \bibinfo {note} {arXiv:1210.0989} (\bibinfo {year} {2012})\BibitemShut {NoStop}%
\bibitem [{\citenamefont {Holtzman}\ and\ \citenamefont
  {Juanes}(2010)}]{Holtzman10}%
  \BibitemOpen
  \bibfield  {author} {\bibinfo {author} {\bibfnamefont {R.}~\bibnamefont
  {Holtzman}}\ and\ \bibinfo {author} {\bibfnamefont {R.}~\bibnamefont
  {Juanes}},\ }\href@noop {} {\bibfield  {journal} {\bibinfo  {journal} {Phys.
  Rev. E}\ }\textbf {\bibinfo {volume} {82}},\ \bibinfo {pages} {046305}
  (\bibinfo {year} {2010})}\BibitemShut {NoStop}%
\bibitem [{\citenamefont {Holtzman}\ \emph {et~al.}(2012)\citenamefont
  {Holtzman}, \citenamefont {Szulczewski},\ and\ \citenamefont
  {Juanes}}]{Holtzman12}%
  \BibitemOpen
  \bibfield  {author} {\bibinfo {author} {\bibfnamefont {R.}~\bibnamefont
  {Holtzman}}, \bibinfo {author} {\bibfnamefont {M.~L.}\ \bibnamefont
  {Szulczewski}}, \ and\ \bibinfo {author} {\bibfnamefont {R.}~\bibnamefont
  {Juanes}},\ }\href@noop {} {\bibfield  {journal} {\bibinfo  {journal} {Phys.
  Rev. Lett.}\ }\textbf {\bibinfo {volume} {108}},\ \bibinfo {pages} {264504}
  (\bibinfo {year} {2012})}\BibitemShut {NoStop}%
\bibitem [{\citenamefont {Kitsunezaki}(2011)}]{Kitsunezaki11}%
  \BibitemOpen
  \bibfield  {author} {\bibinfo {author} {\bibfnamefont {S.}~\bibnamefont
  {Kitsunezaki}},\ }\href@noop {} {\bibfield  {journal} {\bibinfo  {journal}
  {Adv. Powder Technol.}\ }\textbf {\bibinfo {volume} {22}},\ \bibinfo
  {pages} {311} (\bibinfo {year} {2011})}\BibitemShut {NoStop}%
\bibitem [{\citenamefont {Lawn}(1993)}]{Lawn93}%
  \BibitemOpen
  \bibfield  {author} {\bibinfo {author} {\bibfnamefont {B.}~\bibnamefont
  {Lawn}},\ }\href@noop {} {\emph {\bibinfo {title} {Fracture of Brittle
  Solids}}},\ \bibinfo {edition} {2nd}\ ed.\ (\bibinfo  {publisher} {Cambridge
  University Press},\ \bibinfo {address} {Cambridge},\ \bibinfo {year}
  {1993})\BibitemShut {NoStop}%
\bibitem [{\citenamefont {Jr}(1992)}]{Beuth92}%
  \BibitemOpen
  \bibfield  {author} {\bibinfo {author} {\bibfnamefont {J.~L.}\ \bibnamefont
  {Beuth Jr}},\ }\href@noop {} {\bibfield  {journal} {\bibinfo  {journal} {Int. J.
  Solids Structures}\ }\textbf {\bibinfo {volume} {29}},\ \bibinfo {pages}
  {1657} (\bibinfo {year} {1992})}\BibitemShut {NoStop}%
\bibitem [{\citenamefont {Kendall}\ \emph {et~al.}(1987)\citenamefont
  {Kendall}, \citenamefont {Alford},\ and\ \citenamefont
  {Birchall}}]{Kendall87}%
  \BibitemOpen
  \bibfield  {author} {\bibinfo {author} {\bibfnamefont {K.}~\bibnamefont
  {Kendall}}, \bibinfo {author} {\bibfnamefont {N.~M.}\ \bibnamefont {Alford}},
  \ and\ \bibinfo {author} {\bibfnamefont {J.~D.}\ \bibnamefont {Birchall}},\
  }\href@noop {} {\bibfield  {journal} {\bibinfo  {journal} {Proc. R. Soc.
  Lond. A}\ }\textbf {\bibinfo {volume} {412}},\ \bibinfo {pages} {269}
  (\bibinfo {year} {1987})}\BibitemShut {NoStop}%
\end{thebibliography}

%

\end{document}